%% file: main.tex
\documentclass[letterpaper,nomarginnotes,nonarrowgutter]{jpaper}

\usepackage[sort,compress]{cite}
\usepackage{amsmath,amssymb,amsfonts}
\usepackage{algorithmic}
\usepackage{graphicx}
\usepackage{textcomp}
\usepackage[x11names]{xcolor}
\usepackage{booktabs}
\usepackage{multirow}
\usepackage{caption}
\usepackage{footmisc}
\usepackage{fancyhdr}
\usepackage{setspace}
\input{preamble/packages}
\input{preamble/macro}
\usepackage[bookmarks=true,breaklinks=true,letterpaper=true,colorlinks,citecolor=blue,linkcolor=blue,urlcolor=blue]{hyperref}

\newif\iffootnoterule

\makeatletter
\AtBeginDocument{%
\let\latex@@footnoterule\footnoterule

\renewcommand\footnoterule{%
  \iffootnoterule
  \latex@@footnoterule%
  \else
  \advance\skip\footins 4\p@\@plus2\p@\relax%
  \fi
}
}
\makeatother

\makeatletter
\g@addto@macro{\normalsize}{%
  \setlength{\abovedisplayskip}{4pt plus 0.5pt minus 1pt}
  \setlength{\belowdisplayskip}{3pt plus 0.5pt minus 1pt}
  \setlength{\abovedisplayshortskip}{0pt}
  \setlength{\belowdisplayshortskip}{0pt}
  \setlength{\intextsep}{5pt plus 1pt minus 1pt}
  \setlength{\textfloatsep}{3pt plus 1pt minus 1pt}
  \setlength{\skip\footins}{5pt plus 1pt minus 1pt}
  \setlength{\abovecaptionskip}{2pt plus 0pt minus 0pt}}
\makeatother

\usepackage{titlesec}
\titlespacing\section{0pt}{5pt plus 1pt minus 1pt}{2pt plus 1pt minus 1pt}
\titlespacing\subsection{0pt}{5pt plus 1pt minus 1pt}{2pt plus 1pt minus 1pt}
\titlespacing\subsubsection{0pt}{5pt plus 1pt minus 1pt}{2pt plus 1pt minus 1pt}

\newcommand{\thetitle}{DejaVu: Why You Should Write\\to Your DRAM Rows Twice, Carefully}
\title{{\huge\thetitle}}

\author{
Haocong Luo$^{1}$\quad
{\.I}smail~Emir~Y{\"u}ksel$^{1}$\quad
Ataberk Olgun$^{1}$\\[0.3em]
Nisa Bostanci$^{1}$\quad
Orhun Ecemi{\c{s}}$^{2}$\quad
Abdullah Giray Yaglikci$^{3}$\quad
Onur Mutlu$^{1}$\\[0.4em]
{\textit{$^{1}$ETH Zurich\quad
$^{2}$TOBB ET{\"U}\quad
$^{3}$CISPA}}
}

\begin{document}
\maketitle
\thispagestyle{fancy}

\input{sections/00-abstract}
\input{sections/01-introduction}

\footnoteruletrue

\input{sections/02-background}
\input{sections/03-methodology}
\input{sections/04-foundational_results}
\input{sections/05-sensitivity-to-tWR}
\input{sections/06-PUD}
\input{sections/08-Implications}
\input{sections/08a-mitigation}
\input{sections/08c-related_works}

\input{sections/09-conclusion}

\section*{Acknowledgments}

We thank the anonymous reviewers and artifact evaluators of ISCA 2026 for feedback.
We thank the SAFARI Research Group members for their constructive feedback and for providing a stimulating intellectual \arxivaccept{and scientific} environment. 
We acknowledge the generous gift funding provided by our industrial partners (especially Google, Huawei, Intel, Microsoft), which has been instrumental in enabling the research we have been conducting on read disturbance in DRAM in particular and memory systems in general~\cite{mutlu2017rowhammer,mutlu2019processing,mutlu2019rowhammer,mutlu2022modern,mutlu2023fundamentally,mutlu2023retrospective,mutlu2014research,mutlu2023retrospective-raidr,mutlu2023retrospectiveexperimentalstudydata, mutlu2025memory,mutlu2013memory, kakolyris2026columnkeeper}.
This work was in part supported by a Google Security and Privacy Research Award and the Microsoft Swiss Joint Research Center.

\input{main.bbl}
\clearpage
\input{sections/10-artifact-appendix}

\end{document}

%% file: preamble/packages.tex
\usepackage{xspace}
\usepackage{adjustbox}
\usepackage{pdflscape}
\usepackage{listings}
\usepackage[most]{tcolorbox}

%% file: preamble/macro.tex
\newcommand{\arxivaccept}[1]{\iffalse\else#1\fi}
\definecolor{bg}{rgb}{0.95,0.95,0.95}
\definecolor{commentred}{HTML}{E06666}
\definecolor{stringblue}{HTML}{256DCC}
\definecolor{linenumbergreen}{HTML}{2E8B57}
\lstdefinestyle{pythoncompact}{
  language=Python,
  columns=fullflexible,
  basicstyle=\ttfamily\scriptsize,
  numbers=left,
  numberstyle=\tiny\color{linenumbergreen},
  stepnumber=1,
  numbersep=4pt,
  backgroundcolor=\color{bg},
  frame=single,
  framesep=3mm,
  breaklines=true,
  breakindent=0pt,
  postbreak={},
  commentstyle=\itshape\color{commentred},
  stringstyle=\color{stringblue},
  keywordstyle=\bfseries,
  showstringspaces=false,
  tabsize=4,
  xleftmargin=4pt,
  framexleftmargin=8pt,
}

\newcommand{\acmin}[0]{$AC_{min}$\xspace}

\tcbset{before skip=6pt, after skip=6pt}
\definecolor{obscolor}{HTML}{66C2A5}
\definecolor{takecolor}{HTML}{F4A261}

\newtcolorbox[auto counter]{obsx}[3][]{%
    colframe = #2!45,
    colback  = #2!10,
    coltitle = #2!20!black,
    colbacktitle=#2!20,
    coltitle=black,
    fonttitle=\bfseries,
    title=#3~\thetcbcounter.\ ,
    enhanced,
    attach boxed title to top left={yshift=-2.8mm, xshift=0.15cm},
    bottom=-2.2pt,
    #1%
}

\newtcolorbox[auto counter]{tkx}[2][]{%
  enhanced, center title,
  colframe = #2!55!black,
  colback  = #2!10,
  colbacktitle = #2!25,
  left=0.5pt,
  right=0.5pt,
  bottom=2pt,
  top=0.25pt,
  #1%
}

\newcounter{obs}
\newcommand\observation[1]{%
  \refstepcounter{obs}%
  \begin{tkx}{obscolor}%
    \noindent\textbf{Observation~\theobs.} #1%
  \end{tkx}%
}

\newcounter{tkw}
\newcommand\takeaway[1]{%
  \stepcounter{tkw}%
  \begin{tkx}{takecolor}%
    \noindent\textbf{Takeaway~\thetkw.} #1%
  \end{tkx}%
}

\newcommand{\figref}[1]{Figure~\ref{#1}}
\newcommand{\figsref}[1]{Figures~\ref{#1}}
\newcommand{\secref}[1]{Section~\ref{#1}}
\newcommand{\tabref}[1]{Table~\ref{#1}}

\newcommand{\rhmemisolationrefs}[0]{\cite{fournaris2017exploiting, poddebniak2018attacking, tatar2018throwhammer, carre2018openssl, barenghi2018software, zhang2018triggering, bhattacharya2018advanced, google-project-zero, kim2014flipping, rowhammergithub, seaborn2015exploiting, van2016drammer, gruss2016rowhammer, razavi2016flip, pessl2016drama, xiao2016one, bosman2016dedup, bhattacharya2016curious, burleson2016invited, qiao2016new, brasser2017can, jang2017sgx, aga2017good, mutlu2017rowhammer, tatar2018defeating, gruss2018another, lipp2018nethammer, van2018guardion, frigo2018grand, cojocar2019eccploit,  ji2019pinpoint, mutlu2019rowhammer, hong2019terminal, kwong2020rambleed, frigo2020trrespass, cojocar2020rowhammer, weissman2020jackhammer, zhang2020pthammer, yao2020deephammer, deridder2021smash, hassan2021utrr, jattke2022blacksmith, tol2022toward, kogler2022half, orosa2022spyhammer, zhang2022implicit, liu2022generating, cohen2022hammerscope, zheng2022trojvit, fahr2022frodo, tobah2022spechammer, rakin2022deepsteal, park2016statistical, park2016experiments,lim2017active, ryu2017overcoming, yun2018study, yang2019trap, walker2021ondramrowhammer, kim2020revisiting, orosa2021deeper, yaglikci2022understanding, khan2018analysis, agarwal2018rowhammer, li2014write, ni2018write, genssler2022reliability, mutlu2023fundamentally, meyer2026phoenix, Lin2025GPUHammer, aydin2022cyber, mus2022jolt, wang2022research, lefforge2023reverse, fahr2022effects, kaur2022work, cai2022feasibility, li2022cyberradar, roohi2022efficient, staudigl2022neurohammer, yang2022socially, islam2022signature}}

\newcommand{\mitigatingRowHammerAllCitations}[0]{\cite{AppleRefInc, rh-hp,rh-lenovo,greenfield2012throttling, kim2014flipping, kim2014architectural, bains14d, bains14c, bains14a, bains14b, aweke2016anvil, bains2015row, bains2016row, seyedzadeh2017counter, bains2016distributed, son2017making, seyedzadeh2018cbt,irazoqui2016mascat, you2019mrloc, lee2019twice, park2020graphene, yaglikci2021security, yaglikci2021blockhammer, frigo2020trrespass, kang2020cattwo, hassan2021utrr, qureshi2022hydra, saileshwar2022randomized, brasser2017can, konoth2018zebram, van2018guardion, vig2018rapid,  kim2022mithril, lee2021cryoguard, marazzi2022protrr, zhang2022softtrr, joardar2022learning, juffinger2023csi, yaglikci2022hira, saxena2022aqua, enomoto2022efficient, manzhosov2022revisiting, ajorpaz2022evax, naseredini2022alarm, joardar2022machine, tomita2022extracting, zhang2020leveraging,loughlin2021stop, devaux2021method, fakhrzadehgan2022safeguard, saroiu2022price, loughlin2022moesiprime, han2021surround, mutlu2023fundamentally, woo2023scalable, marazzi2023rega, bock2019riprh, wang2021discreet, bennett2021panopticon, olgun2024abacus, bostanci2024comet, canpolat2024understanding, saxena2024start, saxena2024rubix, qureshi2024mint, saxena2024impress, kim2023ddr5, canpolat2025chronus, canpolat2024breakhammer, tugrul2025understanding,yaglikci2024svard,taneja2025dream,vittal2025mopac,lin2025cnc,qazi2025drfm,lu2025counterpoint,qureshi2025autorfm,woo2025dapper,woo2025qprac,bostanci2025understanding, kakolyris2026columnkeeper}}

%% file: sections/00-abstract.tex
\begin{abstract}
\arxivaccept{We} provide the first experimental demonstration of DejaVu, a phenomenon where the data \emph{previously} written to DRAM cells affects DRAM's vulnerability to read disturbance. Our experimental characterization using 112 commercial-off-the-shelf DDR4 DRAM chips from all three major manufacturers shows that, compared to the baseline where we initialize the victim row by writing to it only once, 1) initializing the victim row by \emph{overwriting} the victim row with the opposite data (compared to what was previously written) reduces \acmin (the minimum aggressor row activation count to induce at least one bitflip\arxivaccept{;} a lower \acmin means higher vulnerability to read disturbance), and 2) initializing the victim row by writing the \emph{same} data twice increases \acmin.

We provide two hypotheses to explain DejaVu. First, we hypothesize that overwriting the victim row \arxivaccept{with the opposite data values} causes the under-restoration of charge in the DRAM cells. Second, we hypothesize that the process of overwriting the victim row changes the charge trap states in the active region, affecting the read-disturbance-induced cell leakage current. We conduct controlled experimental characterization to provide \arxivaccept{insight into} these two hypotheses.

To further investigate DejaVu's potential impact on the current passing capability of the DRAM cell access transistors, we characterize the reliability of Processing-Using-DRAM (PUD) operations with DRAM rows initialized with DejaVu patterns. Our experimental characterization of 32-row MAJ-3 operation shows that by overwriting the DRAM rows used in the operation, the number of bitlines that fail to reliably perform MAJ-3 reduces by 32.7\% on average compared to the baseline where the rows are written only once. We hypothesize that DejaVu's effects make the distribution of the current passing capabilities of the access transistors of DRAM cells more uniform compared to the baseline.

Based on our observations, we \arxivaccept{describe} two major implications of DejaVu. We provide an example of how DRAM testing and characterization methodologies should take DejaVu into consideration to 1) accurately characterize the read disturbance vulnerability of DRAM rows under fixed data patterns, and 2) rigorously study the effect of different data patterns on read disturbance \arxivaccept{by avoiding} unintended interference from DejaVu. We also evaluate the additional performance overhead of read disturbance mitigation techniques when their read disturbance thresholds need to be lowered to be secure against DejaVu {and show that they induce higher performance overheads \arxivaccept{(e.g., 6.3\% performance overhead when reducing the read disturbance threshold by 20\% \arxivaccept{as a guardband} to mitigate DejaVu)}}.
\end{abstract}

%% file: sections/01-introduction.tex
\section{Introduction}
\label{sec:intro}
Modern DRAM chips are vulnerable to \emph{read disturbance} \arxivaccept{phenomena, mainly} RowHammer~\cite{kim2014flipping, mutlu2017rowhammer, mutlu2019rowhammer, kim2020revisiting, orosa2021deeper} and RowPress~\cite{luo2023rowpress, luo2024experimental, luo2024rowpressTopPicks}. Repeatedly accessing DRAM rows (i.e., aggressor rows) many times (RowHammer) or keeping the aggressor DRAM rows open for a long period of time (RowPress) induces bitflips in \emph{unaccessed} victim DRAM rows \arxivaccept{that are} physically near the aggressor rows. DRAM read disturbance is a severe threat to system robustness because it breaks the fundamental security principle of memory isolation~\rhmemisolationrefs{}. To design safe, secure, and reliable systems, it is important to comprehensively and rigorously understand and characterize DRAM read disturbance.

\begin{figure}[ht]
    \centering
    \includegraphics[width=\linewidth]{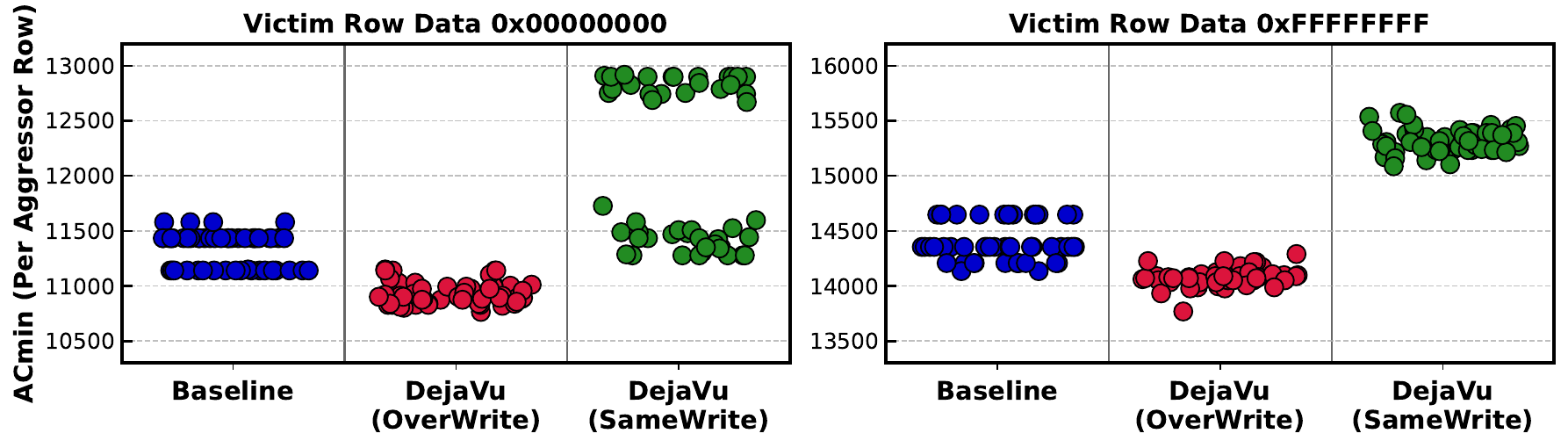}
    \caption{Double-Sided RowHammer ($50^\circ\mathrm{C}$) \acmin distribution across 50 iterations for different victim row initialization methods from one example DRAM row in Mfr. S 8Gb D-Die.}
    \label{fig:acmin_example}
\end{figure}

Prior works in characterizing DRAM read disturbance do \emph{not} study \arxivaccept{how the initialization of the victim DRAM row affects} DRAM read disturbance. In this paper, we provide the first experimental demonstration of DejaVu, a phenomenon where the data \emph{previously} written to DRAM cells affects DRAM's vulnerability to read disturbance. \figref{fig:acmin_example} illustrates an example of how DejaVu changes the minimum aggressor row activation count to induce at least one bitflip (i.e., \acmin, a key metric in measuring a DRAM chip's vulnerability to read disturbance~\cite{luo2023rowpress}) of Double-Sided RowHammer~\cite{kim2014flipping, kim2020revisiting, luo2023rowpress}. The blue dots show the \acmin distribution of the baseline case where we initialize the victim DRAM row with data $X$ by writing to the row \emph{just once}. The red dots show the \acmin distribution where we first write the inverted data $\bar{X}$ to the victim row before \emph{overwriting} it with data $X$ \arxivaccept{(i.e., \emph{OverWrite})}. We observe that \emph{fewer} aggressor row activations are needed to induce the first bitflip in the victim row com{pa}red to the baseline. If we instead write \emph{the same} data $X$ to the victim row twice (\arxivaccept{i.e., \emph{SameWrite},} green dots), \emph{more} aggressor row activations are needed to induce the first bitflip in the victim row com{pa}red to the baseline.

We perform comprehensive experimental characterization of DejaVu on 112 commercial-off-the-shelf DDR4 DRAM chips (14 DIMM modules) from all three major manufacturers spanning a wide range of DRAM die densities and die revisions \arxivaccept{using the FPGA-based DRAM Bender infrastructure~\cite{hassan2017softmc,softmc-safarigithub, olgun2023dram, safari-drambender}}. Our results show that, for double-sided RowHammer, at $50^\circ\mathrm{C}$ compared to the baseline \arxivaccept{where we write to the victim row only once}, 1) overwriting the victim row \arxivaccept{that is \emph{previously} written with data \texttt{0xFF}} with \arxivaccept{data \texttt{0x00}} reduces \acmin by 2.8\% on average (up to 28.1\%) across all tested DRAM rows, 2) overwriting the victim row \arxivaccept{(previously written \texttt{0x00})} with \arxivaccept{data \texttt{0xFF}} reduces \acmin by 3.5\% on average (up to 27.1\%), 3) writing \arxivaccept{\emph{the same} data \texttt{0x00}} to the victim row twice increases \acmin by 3.0\% on average (up to 23.8\%), and 4) writing \arxivaccept{data \texttt{0xFF}} to the victim row twice increases \acmin by 3.6\% on average (up to 46.1\%). \arxivaccept{These three changes and their ``direction'' (i.e., \arxivaccept{whether they reduce or increase} \acmin) are consistent across all tested DRAM chips and tested rows.}

We also find that DejaVu affects the number of DRAM retention failure bitflips. Our characterization results show that initializing the tested row by \arxivaccept{first writing data \texttt{0x00} and then} overwriting with \arxivaccept{data \texttt{0xFF}} induces 10.4\% more (up to 36.7\%) retention failure bitflips compared to writing \arxivaccept{\emph{the same} data \texttt{0xFF}} to the tested row twice.

We provide two hypotheses to explain DejaVu. First, we hypothesize that overwriting the data in the victim row with the opposite \arxivaccept{data values} causes under-restoration of the charge levels in the DRAM cells. Second, we hypothesize that writing different values to the victim row changes the charge trap states in the active region, causing changes in the trap-induced read disturbance leakage~\cite{yang2019trap, Jie2024Understanding, Zhou2024Unveiling, Zhou2024Understanding, Zhou2023Double} \arxivaccept{and the current passing capabilities of the DRAM cell access transistors}. Although we cannot fully verify either of these hypotheses due to limited observability at the \arxivaccept{real} DRAM-chip level, we conduct sensitivity studies \arxivaccept{on real DDR4 DRAM chips using the FPGA-based DRAM Bender infrastructure~\cite{hassan2017softmc,softmc-safarigithub, olgun2023dram, safari-drambender}} to \arxivaccept{gain chip-level insights into these hypotheses.}

\arxivaccept{First, we characterize how \acmin changes as we increase the write recovery time (i.e., we wait longer after the writing to the victim row before precharging the bank) for the second write to the victim DRAM row. We find that 1) the \acmin of SameWrite remains almost unchanged as the write recovery time of the second write increases, and 2) the second write of OverWrite needs a write recovery time more than $400\times$ of the JEDEC standard value~\cite{jedec2017ddr4} to achieve an \acmin distribution close to \arxivaccept{that of} SameWrite. We conclude that the charge under-restoration hypothesis \arxivaccept{(i.e., the first hypothesis described above) alone does \emph{not fully}} explain DejaVu.}

\arxivaccept{Second,} to further investigate DejaVu's potential impact on the current passing capability of the DRAM cell access transistors, we characterize the reliability of Processing-Using-DRAM (PUD)~\cite{seshadri2017ambit, hajinazar2021simdram, yuksel2024simultaneous, yuksel2024functionally, mutlu2024memory} operations with DRAM rows initialized with DejaVu patterns. PUD operations are highly sensitive to the current passing capability of the DRAM cell access transistors because PUD operations rely on a delicate charge sharing process involving \arxivaccept{the simultaneous activation of} multiple input DRAM rows. We find that by leveraging DejaVu, the number of bitlines that fail to \emph{always} successfully perform MAJ-3 operation over 1000 repetitions significantly reduces compared to the baseline case where the DRAM rows are \arxivaccept{written only once}. We observe that, using random data patterns \arxivaccept{that both 1) match real program input data better than all \texttt{0x00}s and all \texttt{0xFF}s, and 2) maximize inter-bitline interference during simultaneous multi-row activation}, when we overwrite the DRAM rows that are previously written with the opposite data values, the number of failed bitlines reduces by 32.7\% (10.7\%) on average compared to the baseline for 32-row (16-row) MAJ-3 operations. When we write the same data twice, the number of failed bitlines reduces by 30.6\% (5.8\%) on average compared to the baseline for 32-row (16-row) MAJ-3 operations. \arxivaccept{We hypothesize that DejaVu's effects at the device-level make the distribution of the current passing capabilities of the access transistors of DRAM cells more uniform, which improves the reliability of PUD operations, compared to the baseline where the DRAM rows are written only once.}

\noindent\textbf{Implications.} Based on our observations, we discuss the two major implications of DejaVu on DRAM read disturbance testing and characterization methodology. First, to accurately characterize \arxivaccept{DRAM's vulnerability to read disturbance (e.g., \acmin)} given a fixed data pattern, the tester should always initialize the victim row by \arxivaccept{first writing the \emph{opposite} data values and then} overwriting it with \arxivaccept{the intended data values (i.e., the OverWrite pattern)} to cover the reduction in \acmin caused by DejaVu. Second, to explore the effect of different data patterns on DRAM read disturbance, the tester should \arxivaccept{avoid accidentally inducing DejaVu effects when initializing the DRAM rows involved in a test. For example, we recommend that testers \emph{always} initialize the victim row by writing the same data values twice (i.e., SameWrite) to avoid attributing the difference in \acmin caused by DejaVu to the change in data patterns.}

We also evaluate the additional performance overhead of read disturbance mitigation techniques \arxivaccept{(e.g., PARA~\cite{kim2014flipping} and \arxivaccept{PRAC~\cite{jedec2024ddr5, canpolat2024understanding, canpolat2025chronus})}} when their read disturbance thresholds need to be lowered to be secure against DejaVu. {We show that DejaVu-caused reductions in read disturbance thresholds can degrade system performance with existing read disturbance mitigations: \arxivaccept{Our evaluation finds that reducing the read disturbance threshold by 20\% to serve as a guardband to mitigate DejaVu causes a 6.3\% performance overhead on average}.}

We make the following key contributions in \arxivaccept{this} paper:
\begin{itemize}
    \item \arxivaccept{We provide} the first experimental demonstration and characterization of DejaVu, \arxivaccept{i.e., the phenomenon} where the data \emph{previously} written to DRAM cells affects DRAM's vulnerability to read disturbance. \arxivaccept{We also show that DejaVu similarly affects DRAM retention failure bitflips.}
    \item We provide hypotheses and experimental insights into the root causes of DejaVu.
    \item We demonstrate that leveraging DejaVu \arxivaccept{(i.e., by writing to the input DRAM rows twice)} improves the reliability of Processing-Using-DRAM (PUD) operations.
    \item We \arxivaccept{describe the implications of DejaVu. We demonstrate} how DRAM read disturbance testing and characterization methodologies should take DejaVu into account to be more rigorous and comprehensive. \arxivaccept{We also evaluate the performance overhead of read disturbance mitigation techniques when they need to be more conservative to mitigate DejaVu.}
\end{itemize}

%% file: sections/02-background.tex
\section{Background}
\label{sec:background}

\subsection{DRAM Organization}
\label{sec:dram_organization}

\begin{figure}[ht]
    \centering
    \includegraphics[width=\linewidth]{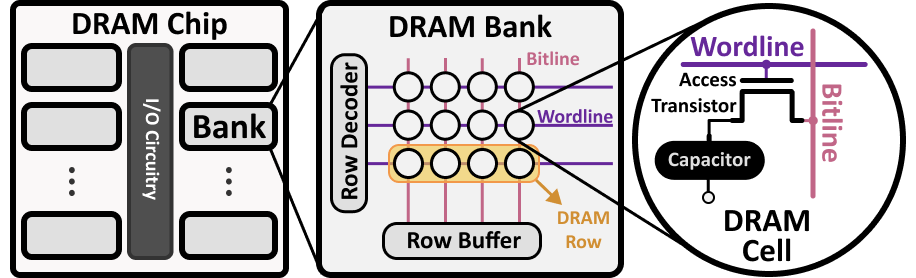}
    \caption{Logical organization of DRAM.}
    \label{fig:dram_organization}
\end{figure}

\figref{fig:dram_organization} shows the logical organization of a DRAM chip. A DRAM chip consists of multiple DRAM banks that \arxivaccept{can operate independently of each other} but share the same I/O resources of the chip. Inside a DRAM bank, DRAM cells are organized into a 2D array. DRAM cells in a row share the same wordline, and DRAM cells in a column share the same bitline that connects the DRAM cell(s) to the row buffer. A DRAM cell consists of a capacitor that stores one bit of information in the form of electric charge and an NMOS access transistor. \arxivaccept{The gate of the access transistor is connected to the wordline of the DRAM row, which controls \arxivaccept{whether or not the capacitor is connected to the bitline.}}

\subsection{DRAM Operation}
\label{sec:dram_operation}
To access DRAM, the DRAM controller first sends a \texttt{PRE} ({i.e., }precharge) command to the bank that closes any opened DRAM row and prepares the bank for the following access. Second, the DRAM controller sends an \texttt{ACT} ({i.e., }activate) command to activate ({i.e., }open) a DRAM row in the bank. To open a DRAM row, the DRAM drives the wordline high to connect the capacitors of the DRAM cells in the row to the bitlines. The capacitor shares its charge with the bitline, creating a voltage disturbance \arxivaccept{the bitline sense amplifiers (BLSAs) in the row buffer can sense and amplify}. After charge sharing, the DRAM row needs to be kept open for some time so that the BLSA can fully restore the charge level in the capacitor of the DRAM cell. When writing data to the DRAM, the memory controller needs to obey the \emph{write recovery} timing constraint between the arrival of write data to the DRAM and sending a \texttt{PRE} command to close the row to allow sufficient charge restoration in the DRAM cells {(i.e., tWR)}.

\subsection{DRAM Read Disturbance}
\label{sec:dram_read_disturbance}
Read disturbance is a phenomenon in DRAM where accessing a DRAM row (aggressor row) causes bitflips in \emph{unaccessed} DRAM rows \arxivaccept{that are} physically nearby (victim row). RowHammer~\cite{kim2014flipping, kim2020revisiting, orosa2021deeper, luo2024experimental, mutlu2019rowhammer, mutlu2017rowhammer, nam2024dramscope} and RowPress~\cite{luo2023rowpress, luo2024experimental, luo2024rowpressTopPicks, nam2024dramscope} are two widespread examples of read disturbance in modern DRAM chips. RowHammer induces bitflips in victim rows by repeatedly opening and closing (i.e., hammering) aggressor row(s) many times. RowPress induces bitflips in victim rows by keeping the aggressor row(s) open for a long period of time without needing as many aggressor row activations as RowHammer.

%% file: sections/03-methodology.tex
\section{DejaVu Characterization Methodology}
\label{sec:methodology}

\subsection{DRAM Characterization Infrastructure}
\label{sec:infra}
\begin{figure}[ht]
    \centering
    \includegraphics[width=0.95\linewidth]{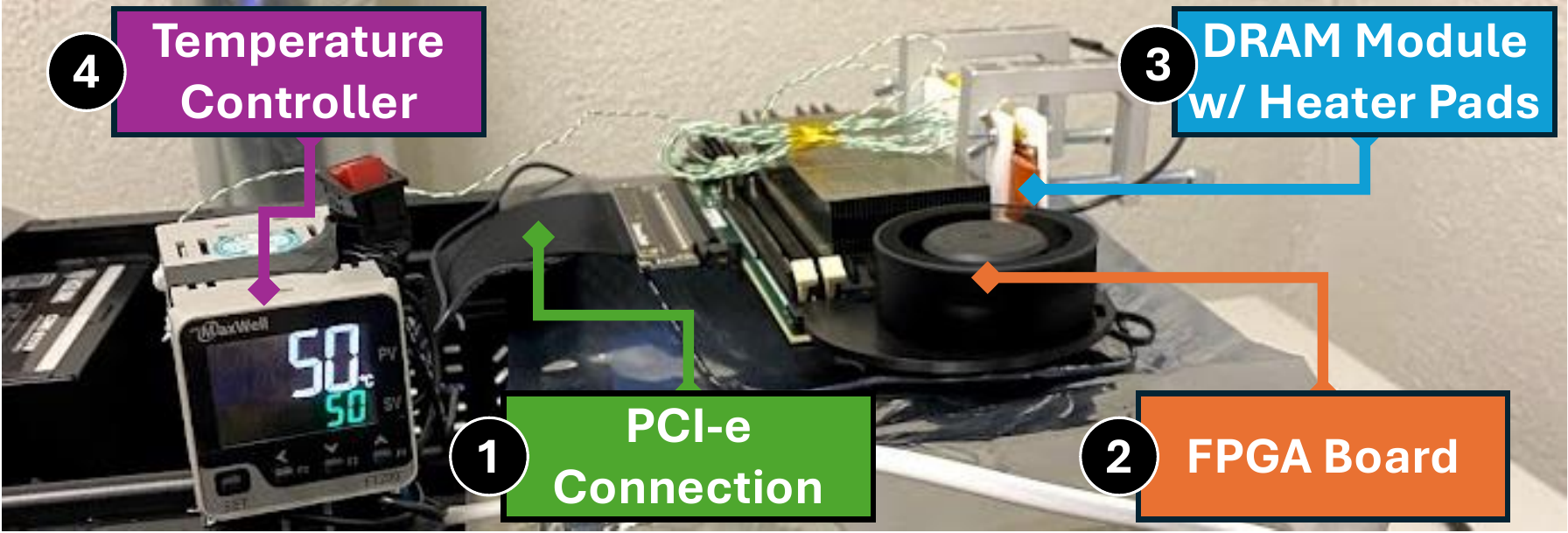}
    \caption{Our DRAM testing infrastructure.}
    \label{fig:dbender}
\end{figure}

\begin{figure}[ht]
    \centering
    \includegraphics[width=0.95\linewidth]{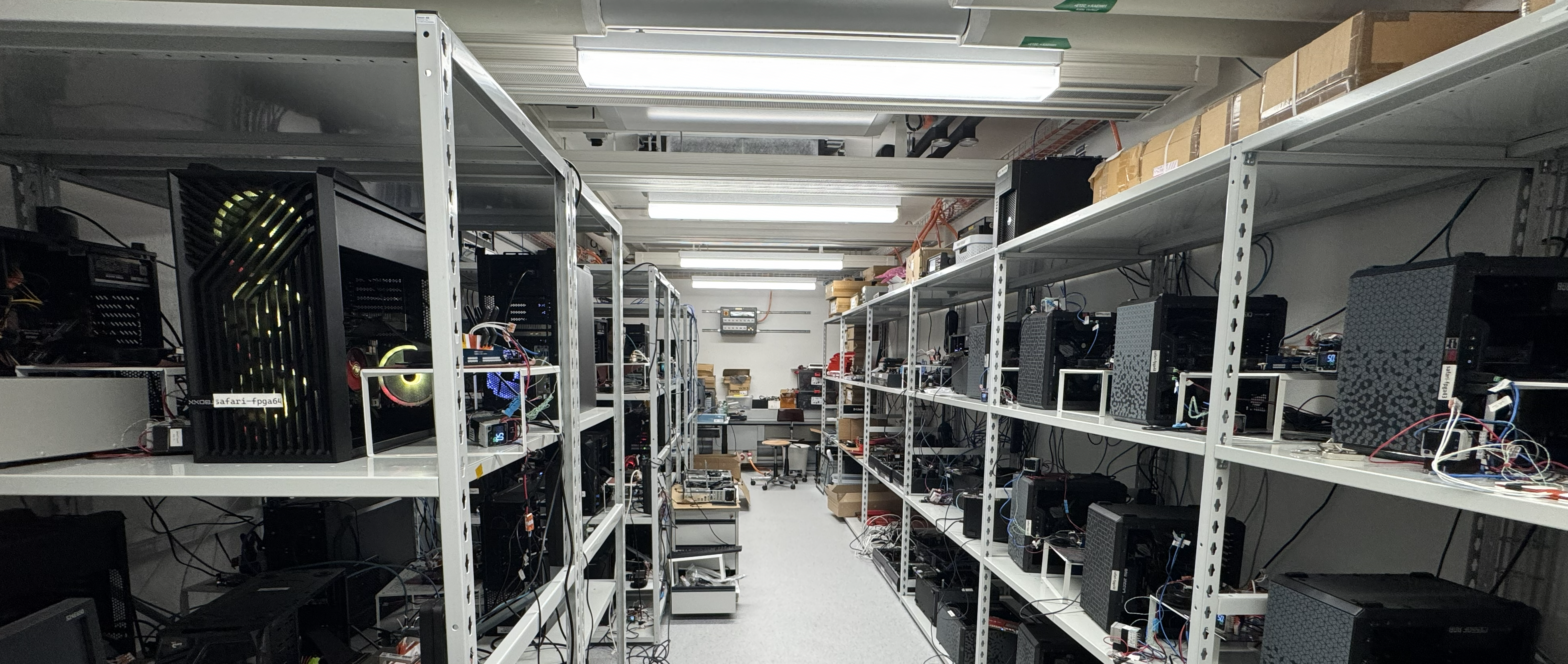}
    \caption{Our lab hosting the testing infrastructure.}
    \label{fig:lab_photo}
\end{figure}

We characterize DejaVu on commercial-off-the-shelf DDR4 DRAM chips using DRAM Bender~\cite{hassan2017softmc,softmc-safarigithub, olgun2023dram, safari-drambender}, an FPGA-based DRAM testing infrastructure that enables direct and fine-grained control of DRAM commands, timings, and temperature.~\figref{fig:dbender} shows our testing infrastructure. ~\figref{fig:lab_photo} shows our lab hosting the infrastructure. 

The FPGA executes test programs that send DRAM commands with precise timings (1.5ns granularity) to the DRAM chips under test. A host PC generates the test programs and collects experiment results. A set of heater pads are attached to the DRAM chips, controlled by a temperature controller that can keep the temperature of the DRAM chips at programmed levels. Our infrastructure disables rank-level ECC so we can directly observe bitflips that happen at the circuit level.

\subsection{DRAM Chips Characterized}
\label{sec:chips}

\tabref{tab:chips} lists the 112 commercial-off-the-shelf DDR4 DRAM chips (14 modules) we characterize in this work. We test DRAM chips from all three major manufacturers (S, H, and M), spanning a wide variety of die densities and revisions. We reverse engineer the internal DRAM row address mapping schemes of all the chips we characterize to precisely place the aggressor and victim rows in our characterization.

\begin{table}[ht]
\centering
\caption{DRAM chips tested.}
\begin{adjustbox}{width=0.98\columnwidth}
\begin{tabular}{@{}c|c|c|c|c|c|c@{}}
\toprule
\textbf{Mfr.} & \textbf{ID} & \textbf{Die Revision} & \textbf{Die Density} & \textbf{DQ} & \textbf{Num. Modules} & \textbf{Num. Chips} \\ \toprule
S & S0    & D & 8 Gb  & x8 & 1 & 8  \\
S & S1    & M & 16 Gb & x8 & 1 & 8  \\
S & S2    & A & 16 Gb & x8 & 1 & 8  \\
S & S3    & B & 16 Gb & x8 & 1 & 8  \\
S & S4    & C & 16 Gb & x8 & 1 & 8  \\ \midrule
H & H0    & A & 8 Gb  & x8 & 1 & 8  \\
H & H1    & C & 8 Gb  & x8 & 1 & 8  \\
H & H2    & D & 8 Gb  & x8 & 1 & 8  \\
H & H3    & A & 16 Gb & x8 & 1 & 8  \\ \midrule
M & M0    & E & 8 Gb  & x8 & 1 & 8  \\
M & M1    & R & 8 Gb  & x8 & 1 & 8  \\
M & M2-M4 & F & 16 Gb & x8 & 3 & 24 \\ \bottomrule
\end{tabular}%
\end{adjustbox}
\label{tab:chips}
\end{table}

We do not test DDR5 or LPDDR5 DRAM chips because 1) there is currently no testing platform available that provides the same degree of low-level and fine-grained control of DRAM commands and timings as DRAM Bender for DDR4, and 2) prior work~\cite{marazzi2024hifidram} shows that there is no fundamental difference in the DRAM cell array between DDR4 and DDR5. Therefore, we believe testing commercial-off-the-shelf DDR4 DRAM chips is enough to reveal the intrinsic \arxivaccept{and previously unreported} DRAM behavior at the circuit-level.

\subsection{True- and Anti-Cell Configurations}
We reverse engineer the true- and anti-cell layout of tested DRAM chips using retention failure analyses, similar to prior works~\cite{patel2017reaper, orosa2021codic, luo2023rowpress, kim2020revisiting, kraft2018improving, nam2024dramscope, luo2025revisiting}, in a best-effort manner. We adjust the data pattern written to the DRAM rows according to the reverse-engineered true- and anti-cell layout.

\subsection{DRAM Access Patterns}
\label{sec:access_pattern}
\begin{figure}[ht]
    \centering
    \includegraphics[width=\linewidth]{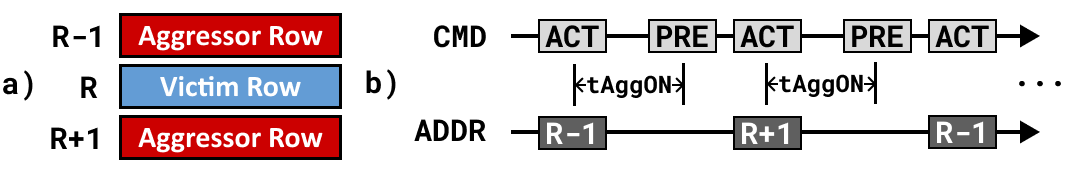}
    \caption{Double-sided RowHammer/RowPress DRAM access pattern.}
    \label{fig:rd_pattern}
\end{figure}

\noindent\textbf{Read Disturbance.} To characterize the read disturbance vulnerability, we use the double-sided access pattern where two aggressor rows sandwich a victim row, as~\figref{fig:rd_pattern} a) shows. In the double-sided pattern, the two aggressors are activated \arxivaccept{in an alternating manner}, as~\figref{fig:rd_pattern} b) shows. For each aggressor row activation, we denote the interval between the corresponding \texttt{ACT} and \texttt{PRE} commands as the aggressor row on time (tAggON\arxivaccept{~\cite{luo2023rowpress}}). When tAggON is the minimum amount of time allowed by the JEDEC standard (we use 36ns), the access pattern is a RowHammer pattern. When tAggON is larger than 36ns, it is a RowPress pattern.

\noindent\textbf{Victim Row Initialization.} The pseudocode in Listing~\ref{lst:row_init} illustrates how we initialize the aggressor and victim rows in the double-sided access pattern differently for the baseline case and the two DejaVu cases (i.e., \emph{OverWrite} and \emph{SameWrite}).

\begin{figure}[tbp]
\centering
\begin{minipage}{0.9\columnwidth}
\begin{lstlisting}[style=pythoncompact,basicstyle=\ttfamily\footnotesize,caption={Pseudocode of the baseline and DejaVu DRAM row initialization procedures.},label={lst:row_init}]
def init_rows(R, aggr_data, victim_data, case):
    # Initialize the aggressor rows with aggr_data
    write_row(R-1, aggr_data)
    write_row(R+1, aggr_data)

    # Initialize the victim row with victim_data
    # Baseline case
    if   case == "Baseline":
        write_row(R, victim_data)
    # DejaVu cases
    else if case == "OverWrite":
        # First write opposite data to victim
        write_row(R, ~victim_data)
        # Then write actual data to victim
        write_row(R, victim_data)
    else if case == "SameWrite":
        # Write the same data to victim twice
        write_row(R, victim_data)
        write_row(R, victim_data)
\end{lstlisting}
\end{minipage}
\end{figure}

\begin{figure}[ht]
    \centering
    \includegraphics[width=\linewidth]{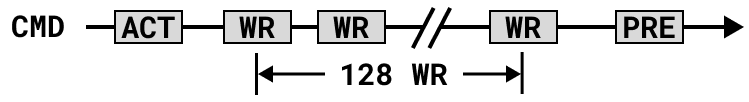}
    \caption{DRAM command sequence of \texttt{write\_row}.}
    \label{fig:wr_cmd_sequence}
\end{figure}

\figref{fig:wr_cmd_sequence} illustrates the DRAM command sequence of \arxivaccept{the \texttt{write\_row} function in Listing 1}. We first activate (ACT) the row, then send 128 write (WR) commands to all 128 cache lines in the DRAM row, and finally precharge (PRE) the bank. We respect all related DRAM timing constraints (\texttt{tRCD}, \texttt{tCCD\_L}, \texttt{tWR}, \texttt{tRP}\arxivaccept{~\cite{kim2012case,lee2013tiered, lee2015aldram}}) when writing to the DRAM row.

\noindent\textbf{Measuring \acmin.} We quantify the DRAM read disturbance vulnerability using the minimum aggressor row activation count \arxivaccept{to induce a bitflip} (\acmin). For a given victim row, we use a bisection-based method to iteratively measure \acmin until the difference in measurements between two consecutive iterations is less than 10. In each iteration, {whether} we find bitflips or not, we always re-initialize the aggressor and victim rows based on the procedure in Listing~\ref{lst:row_init}. 

We strictly control the execution time of each iteration of our \acmin measurement algorithm. We make sure that each iteration takes strictly less than 64ms (i.e., the refresh window of DDR4~\cite{jedec2017ddr4}) to execute, to avoid observing any retention failure bitflips in the \acmin measurement. During each iteration, our testing infrastructure does not issue any auto-refresh commands to make the timings of our testing program precise.

%% file: sections/04-foundational_results.tex
\section{\arxivaccept{Foundational Characterization Results}}
\label{sec:foundational}
To comprehensively characterize DejaVu, we randomly sample 128 victim rows each in all DRAM modules we test. For each victim row, we repeat the \acmin measurement 50 times. We test both \arxivaccept{\texttt{0x00} and \texttt{0xFF} data patterns in the victim DRAM rows. The aggressor rows are always initialized with the \emph{opposite} data pattern compared to the victim row.}

\subsection{RowHammer \acmin Characterization Results}
\label{sec:rh_acmin_results}
\begin{figure}[ht]
    \centering
    \includegraphics[width=\linewidth]{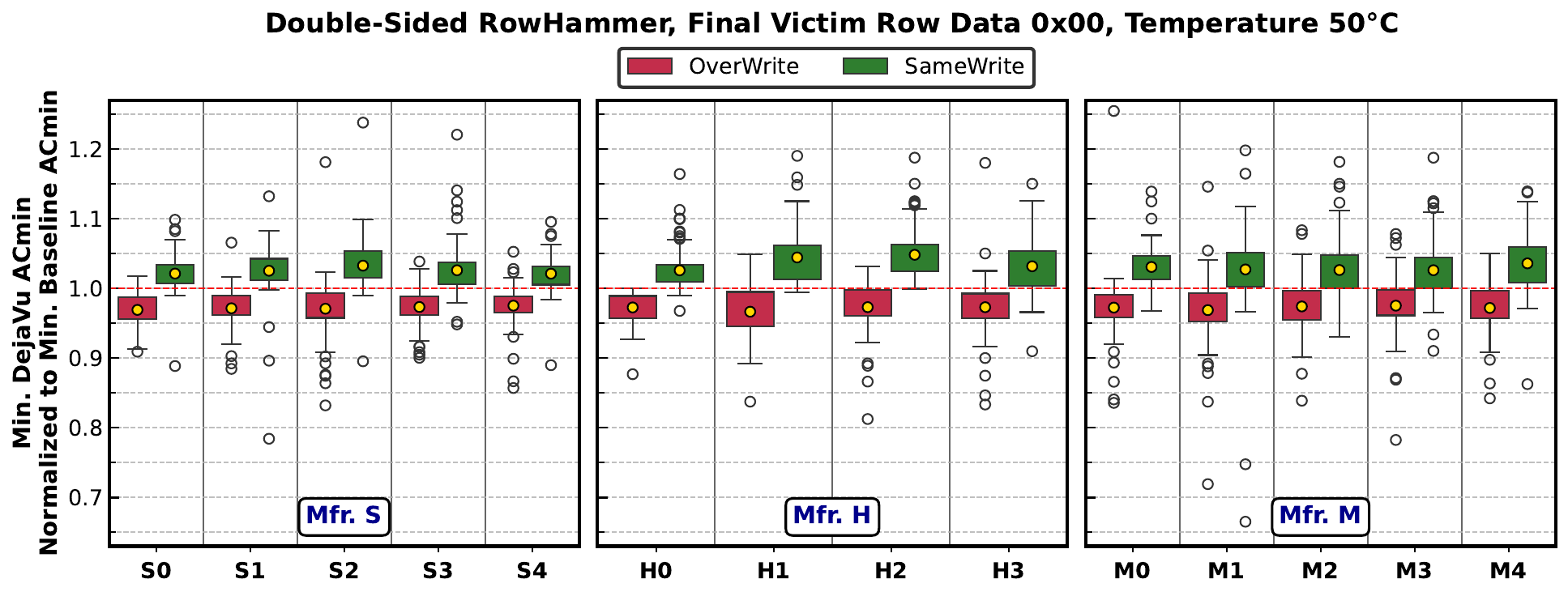}
    \caption{Minimum DejaVu \acmin normalized to minimum baseline \acmin, Double-Sided RowHammer, victim data \texttt{0x00}, $50^\circ\mathrm{C}$. \arxivaccept{Distribution across all 128 tested rows per module.}}
    \label{fig:RH_acmin_min_allrows_00}
\end{figure}

\figref{fig:RH_acmin_min_allrows_00} shows the distribution of the \emph{minimum} DejaVu \acmin (y-axis, measured across all 50 repetitions for each of the tested rows, red for OverWrite, green for SameWrite) normalized to the \emph{minimum} baseline \arxivaccept{\acmin (i.e., victim row written only once)} for Double-Sided RowHammer at a temperature of $50^\circ\mathrm{C}$, for victim data pattern \texttt{0x00}, in box and whisker plots. The box spans the first quartile (Q1) and the third quartile (Q3) of the data. The whiskers span $1.5\times$ the interquartile range (i.e., $1.5\times(Q3-Q1)$) from the Q3 and Q1. Data values outside of the whiskers are outliers (white dots). Note that the yellow dot in the box represents the \emph{geometric} mean of the normalized \acmin values. We make the following three observations from the data.

\observation{DejaVu is a widespread phenomenon across DRAM chips from all three major manufacturers.}

\observation{By overwriting the victim row data, \acmin consistently decreases, while by writing the same data to the victim row twice, \acmin consistently increases.}
We observe that, except for a few outliers, by overwriting the victim row with \texttt{0x00} (i.e., the row is previously written with \texttt{0xFF}\arxivaccept{; the OverWrite pattern}), the minimum \acmin reduces by 2.8\% on average across all tested rows, DRAM chips, and all three manufacturers (2.8\% for Mfr. S, 2.9\% for Mfr. H, 2.8\% for Mfr. M). By writing the same \texttt{0x00} data to the victim row twice \arxivaccept{(i.e., the SameWrite pattern)}, the minimum \acmin increases by 3.0\% on average across all tested rows, DRAM chips, and all three manufacturers (2.5\% for Mfr. S, 2.9\% for Mfr. H, 3.7\% for Mfr. M).

\observation{Although the average decrease (increase) in \acmin is small, there are outliers that significantly change the minimum \acmin measured.}
We observe that certain outliers significantly decrease (for OverWrite) or increase (for SameWrite) the minimum \acmin in certain tested victim rows. For example, the maximum decrease in minimum \acmin caused by OverWrite is 16.8\%, 18.8\%, and 28.1\%, for Mfr. S, H, and M, respectively. The maximum increase in minimum \acmin caused by SameWrite is 23.8\%, 19.0\%, and 19.8\%, for Mfr. S, H, and M, respectively. 

\begin{figure}[ht]
    \centering
    \includegraphics[width=\linewidth]{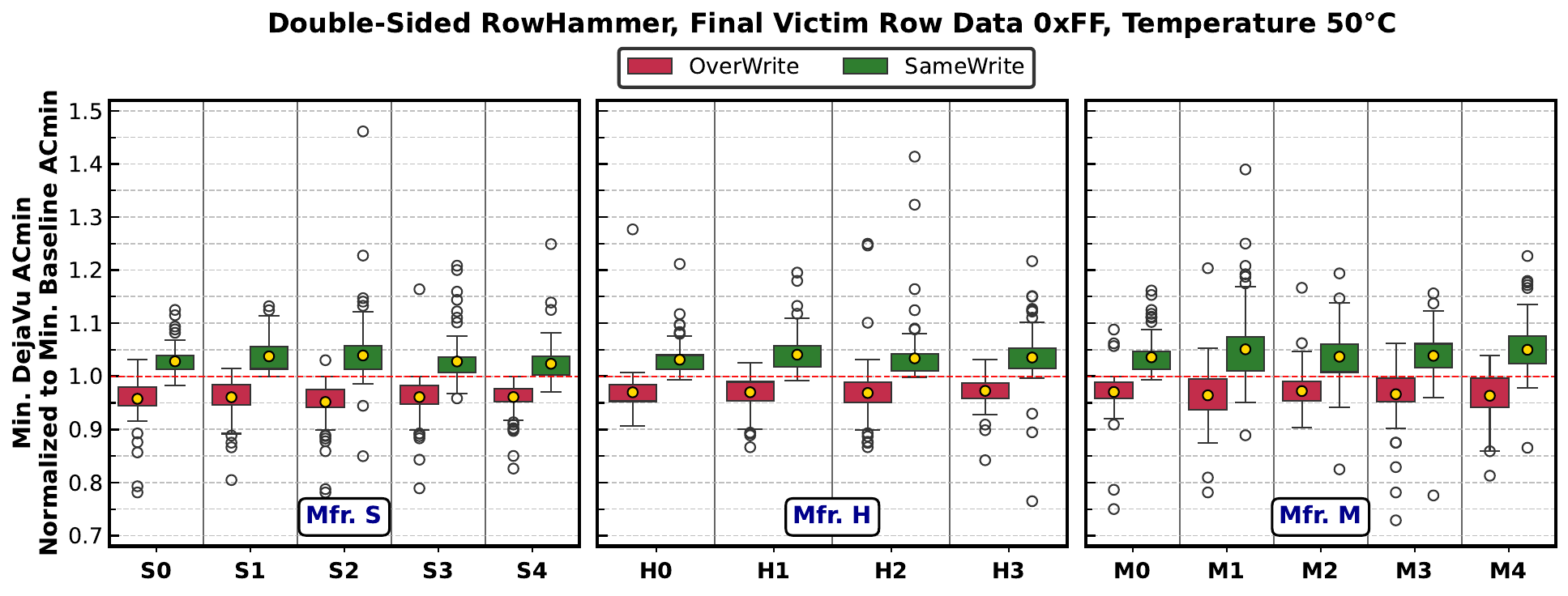}
    \caption{Minimum DejaVu \acmin normalized to minimum baseline \acmin, Double-Sided RowHammer, victim data \texttt{0xFF}, $50^\circ\mathrm{C}$. \arxivaccept{Distribution across all 128 tested rows per module.}}
    \label{fig:RH_acmin_min_allrows_FF}
\end{figure}

\figref{fig:RH_acmin_min_allrows_FF} shows the same distribution as in~\figref{fig:RH_acmin_min_allrows_00}, but with the victim data pattern being \texttt{0xFF}. We make the following observation from the data. 

\observation{The change in \acmin caused by DejaVu is larger when the victim data pattern is \texttt{0xFF} compared to \texttt{0x00}.}
We observe that by overwriting the victim row with \texttt{0xFF} (i.e., the row is previously written with \texttt{0x00}), the minimum \acmin reduces by 3.5\% (compared to 2.8\% with the \texttt{0x00} data pattern) on average across all \arxivaccept{1792} tested rows, \arxivaccept{112} DRAM chips, and all three manufacturers (4.2\% for Mfr. S, 3.0\% for Mfr. H, 3.3\% for Mfr. M, compared to 2.8\%, 2.9\%, and 2.8\%, with the \texttt{0x00} data pattern). By writing the same \texttt{0xFF} data to the victim row twice, the minimum \acmin increases by 3.6\% (compared to 3.0\% with the \texttt{0x00} data pattern) on average across all tested rows, DRAM chips, and all three manufacturers (3.1\% for Mfr. S, 3.5\% for Mfr. H, 4.2\% for Mfr. M, compared to 2.5\%, 2.9\%, and 3.7\%, with the \texttt{0x00} data pattern).

The maximum decrease in minimum \acmin caused by OverWrite is 26.2\%, 15.8\%, and 27.1\%, for Mfr. S, H, and M, respectively (compared to 16.8\%, 18.8\%, and 28.1\% with the \texttt{0x00} data pattern). The maximum increase in minimum \acmin caused by SameWrite is 46.1\%, 41.4\%, and 38.9\%, for Mfr. S, H, and M, respectively (compared to 23.8\%, 19.0\%, and 19.8\% with the \texttt{0x00} data pattern). 

We hypothesize that one of the reasons for DejaVu to cause a larger change in \acmin is that since the DRAM cell access transistor is an NMOS, it is more difficult to restore a ``\texttt{1}'' than a ``\texttt{0}''\arxivaccept{~\cite{luo2020clr,keeth2001dram}}. We provide more characterization, analyses, and hypotheses on the relationship between DejaVu and DRAM cell charge restoration in~\secref{sec:hypotheses} and~\secref{sec:sensitivity}.

\takeaway{DejaVu causes consistent changes in \acmin on DRAM chips from all three manufacturers for Double-Sided RowHammer. \arxivaccept{By overwriting the victim row with data that is opposite to the previously written data, \acmin consistently decreases, while by writing the same data to the victim row twice, \acmin consistently increases.}}
\takeaway{The DejaVu-induced changes in \acmin are larger when writing \texttt{0xFF} to the victim row compared to writing \texttt{0x00} for Double-Sided RowHammer.}

We also characterize Double-Sided RowHammer \acmin of DejaVu at a higher DRAM temperature of $80^\circ\mathrm{C}$. \arxivaccept{\tabref{tab:acmin_temp} shows 1) the changes in the minimum \acmin of DejaVu patterns normalized to the baseline \acmin where the victim row is written only once, and 2) the maximum change in minimum \acmin caused by DejaVu across all tested rows and repetitions compared to the baseline \acmin, for both $50^\circ\mathrm{C}$ and $80^\circ\mathrm{C}$.}
 
\arxivaccept{We make similar observations in how DejaVu changes the minimum \acmin compared to the baseline across different DejaVu write patterns and data patterns at $80^\circ\mathrm{C}$ as for $50^\circ\mathrm{C}$. For example, at both $50^\circ\mathrm{C}$ and $80^\circ\mathrm{C}$, SameWrite consistently increases \acmin while OverWrite consistently decreases \acmin compared to the baseline.}

\begin{table}[ht]
\centering
\resizebox{\columnwidth}{!}{%
\begin{tabular}{@{}c|c||c|cc|cc@{}}
\toprule
\textbf{\begin{tabular}[c]{@{}c@{}}DejaVu\\ Pattern\end{tabular}} &
  \textbf{\begin{tabular}[c]{@{}c@{}}Final\\ Victim \\Data\end{tabular}} &
  \textbf{Mfr.} &
  \textbf{\begin{tabular}[c]{@{}c@{}}Norm. Min.\\ ACmin\\ @ 50°C\end{tabular}} &
  \textbf{\begin{tabular}[c]{@{}c@{}}Norm. Min.\\ ACmin\\ @ 80°C\end{tabular}} &
  \textbf{\begin{tabular}[c]{@{}c@{}}Max. Change\\ in ACmin\\ @ 50°C\end{tabular}} &
  \textbf{\begin{tabular}[c]{@{}c@{}}Max. Change\\ in ACmin\\ @ 80°C\end{tabular}} \\ \toprule
\multirow{6}{*}{SameWrite} & \multirow{3}{*}{\texttt{0x00}} & S & 2.5\%  & 2.1\%  & 23.8\%  & 36.0\%  \\
                           &                       & M & 2.9\%  & 2.2\%  & 19.8\%  & 23.2\%  \\
                           &                       & H & 3.7\%  & 3.2\%  & 19.0\%  & 21.2\%  \\ \cmidrule(lr){2-7}
                           & \multirow{3}{*}{\texttt{0xFF}} & S & 3.1\%  & 2.5\%  & 46.1\%  & 31.8\%  \\
                           &                       & M & 4.2\%  & 3.2\%  & 38.9\%  & 31.5\%  \\
                           &                       & H & 3.5\%  & 3.2\%  & 41.4\%  & 30.7\%  \\ \midrule \midrule
\multirow{6}{*}{OverWrite} & \multirow{3}{*}{\texttt{0x00}} & S & -2.8\% & -2.2\% & -16.8\% & -18.5\% \\
                           &                       & M & -2.8\% & -2.3\% & -28.1\% & -21.9\% \\
                           &                       & H & -2.9\% & -2.4\% & -18.8\% & -18.8\% \\ \cmidrule(lr){2-7}
                           & \multirow{3}{*}{\texttt{0xFF}} & S & -4.2\% & -3.5\% & -26.2\% & -20.3\% \\
                           &                       & M & -3.3\% & -2.9\% & -27.1\% & -42.9\% \\
                           &                       & H & -3.0\% & -3.0\% & -15.8\% & -25.0\% \\ \bottomrule
\end{tabular}%
}
\caption{The changes in normalized minimum \acmin for different DejaVu cases and victim row data patterns at $50^\circ\mathrm{C}$ and $80^\circ\mathrm{C}$. \arxivaccept{Result across all} 1792 tested rows (128 rows per module; 14 modules in total).}
\label{tab:acmin_temp}
\end{table}

\observation{DejaVu changes \acmin in the same direction at a higher temperature of $80^\circ\mathrm{C}$ compared to $50^\circ\mathrm{C}$.}

\subsection{RowPress \acmin Characterization Results}
\begin{figure}[ht]
    \centering
    \includegraphics[width=\linewidth]{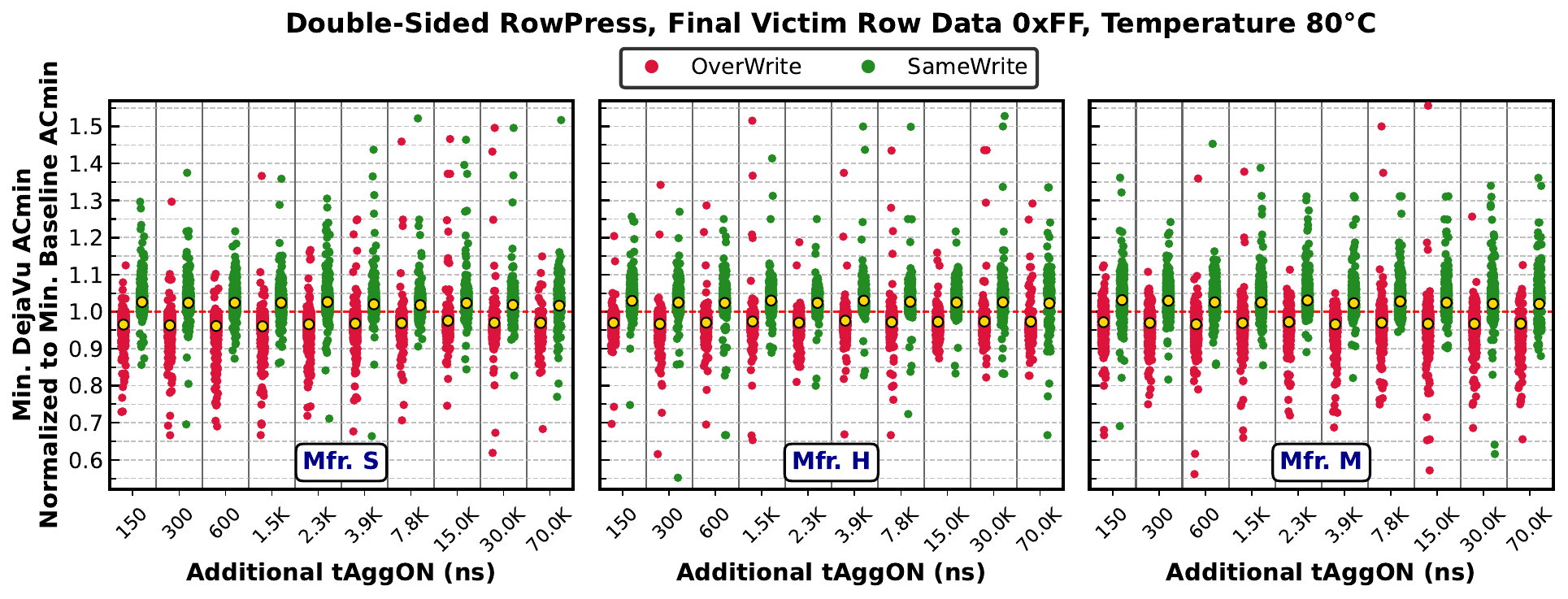}
    \caption{Minimum DejaVu \acmin normalized to minimum baseline \acmin, Double-Sided RowPress, victim data \texttt{0xFF}, $80^\circ\mathrm{C}$. \arxivaccept{Result across all} 1792 tested rows (128 rows per module; 14 modules in total).}
    \label{fig:RP_acmin_min_allrows_FF}
\end{figure}

\figref{fig:RP_acmin_min_allrows_FF} shows the distribution of the \emph{minimum} DejaVu \acmin (y-axis, measured across all 50 repetitions for each of \arxivaccept{all the 1792 tested rows (128 rows per module; 14 modules in total)}, red for OverWrite, green for SameWrite) normalized to the \emph{minimum} baseline \acmin for Double-Sided RowPress with different \emph{additional} tAggON (x-axis) at a temperature of $80^\circ\mathrm{C}$, for victim data pattern \texttt{0xFF}, in jittered scatter plots. The yellow dot represents the geometric mean. We do not show the data in a box and whiskers plot for clarity due to the large number of outliers. We also crop the top of the y-axis to make the majority of the \acmin distribution more readable. We observe that DejaVu causes consistent changes in \acmin on DRAM chips from all three manufacturers for Double-Sided RowPress, similar to Double-Sided RowHammer.

\begin{figure}[ht]
    \centering
    \includegraphics[width=\linewidth]{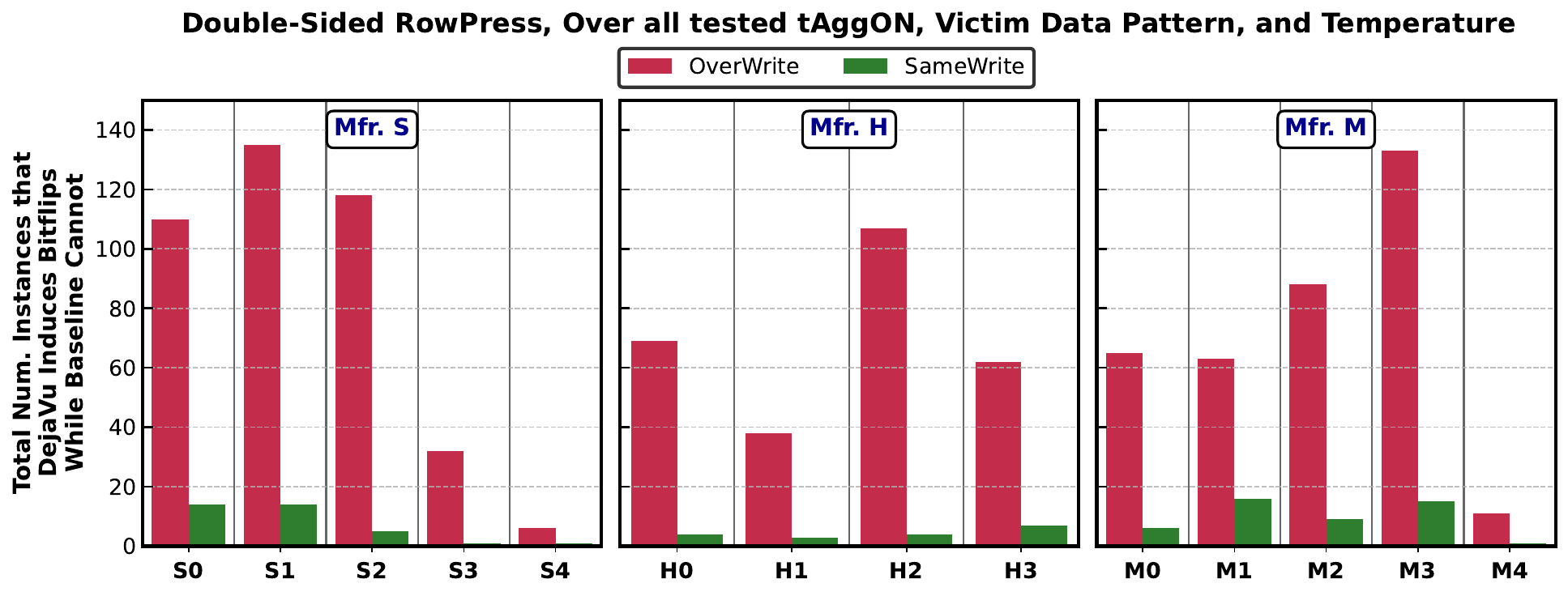}
    \caption{The total number of instances across all tested victim rows, tAggON, victim data pattern, and temperatures that RowPress bitflips are induced \emph{only} when the victim rows are initialized with either of the two DejaVu patterns (OverWrite and SameWrite) but not the baseline pattern in all 50 iterations. \arxivaccept{Summed across all 128 tested rows per module.}}
    \label{fig:RP_new_bitflips}
\end{figure}

\arxivaccept{The \acmin distributions shown in \figref{fig:RP_acmin_min_allrows_FF} do not show the full picture of how DejaVu affects RowPress bitflips because} the plotting methodology cannot illustrate a scenario where the baseline victim row initialization case fails to induce \emph{any} bitflips, but the two DejaVu cases can.~\figref{fig:RP_new_bitflips} shows the total number of instances (i.e., across all tested victim rows, tAggON, \arxivaccept{both \texttt{0x00} and \texttt{0xFF} victim data patterns}, and temperatures) \arxivaccept{in which bitflips are induced} \emph{only} when the victim rows are initialized with either of the \arxivaccept{two DejaVu patterns (OverWrite and SameWrite), \arxivaccept{i.e., the baseline where the victim row is written only once} does not induce \emph{any} bitflips in all 50 iterations.}

\newpage
We observe that by initializing the victim row with DejaVu, we can induce RowPress bitflips in more scenarios compared to the baseline case where we write to the victim only once. Overwriting the victim row data is more effective at inducing these new bitflips compared to writing the same data to the victim row.

\takeaway{DejaVu \arxivaccept{worsens} DRAM's vulnerability to RowPress by inducing bitflips \arxivaccept{that are} not inducible with \arxivaccept{best prior methods}.}

\section{Retention Failure Bitflip Characterization Results}
\label{sec:retention}
We also characterize how DejaVu affects DRAM retention failure bitflips~\cite{liu2013experimental,khan2014efficacy,patel2017reaper}. We perform retention failure tests on the same rows as the victim rows used in the $AC_{min}$ tests. For each row we test, we initialize it with \arxivaccept{final data pattern \texttt{0xFF} using either the OverWrite or SameWrite pattern (as described in Listing 1). For all other rows in the bank, we initialize them with data pattern \texttt{0x00}.} We set the DRAM temperature to $95^\circ\mathrm{C}$ and pause DRAM refresh for 4 seconds to quickly induce a large number of retention failure bitflips~\cite{patel2017reach}. For each row we test, we repeat the experiment 50 times.

\begin{figure}[ht]
    \centering
    \includegraphics[width=0.95\linewidth]{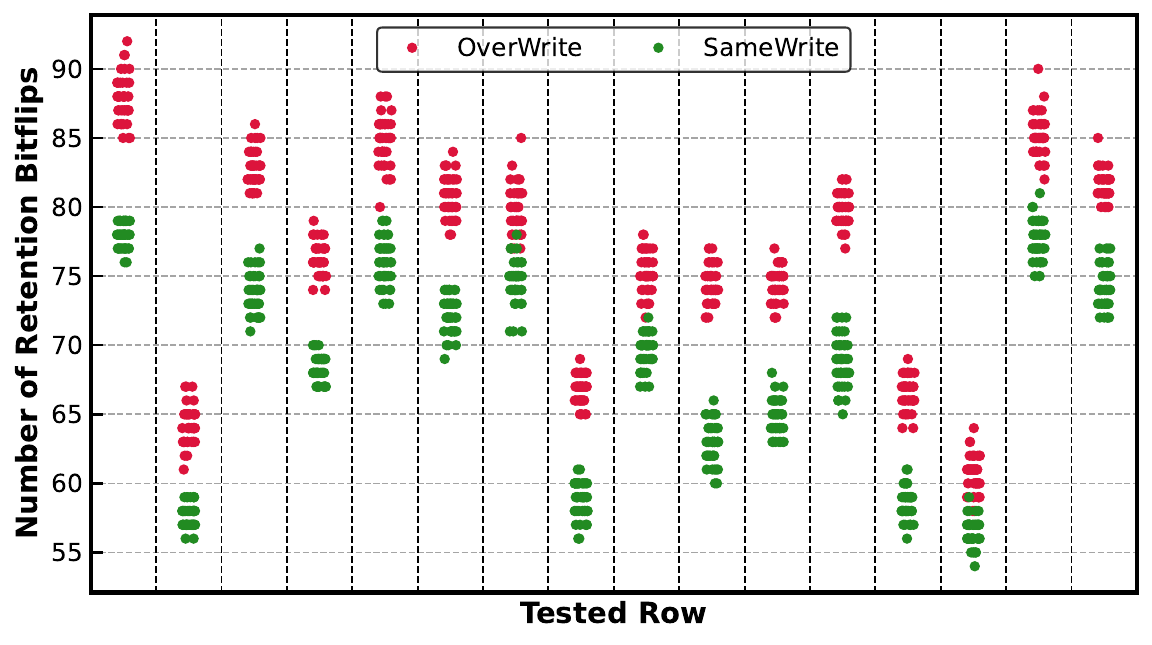}
    \caption{Example distribution of DRAM retention failure bitflips from \arxivaccept{16 example tested rows initialized with} OverWrite and SameWrite from one example module.}
    \label{fig:ret_example}
\end{figure}

\figref{fig:ret_example} shows an example of the distribution of DRAM retention failure bitflips after we initialize the tested row with the OverWrite (red) or SameWrite (green) pattern \arxivaccept{(i.e., for OverWrite, first writing data \texttt{0x00} and then overwriting with data \texttt{0xFF}; for SameWrite, writing data \texttt{0xFF} twice)} across 50 experiment repetitions from one tested module from Mfr. S. We observe that initializing the row with the OverWrite pattern \emph{consistently} induces more retention failure bitflips compared to the SameWrite pattern. The geometric mean of the average number of retention failure bitflips from OverWrite \arxivaccept{is $1.12\times$ (up to $1.24\times$) that from SameWrite across all 128 tested rows and 50 repetitions \arxivaccept{in this example module}.}

\begin{figure}[ht]
    \centering
    \includegraphics[width=\linewidth]{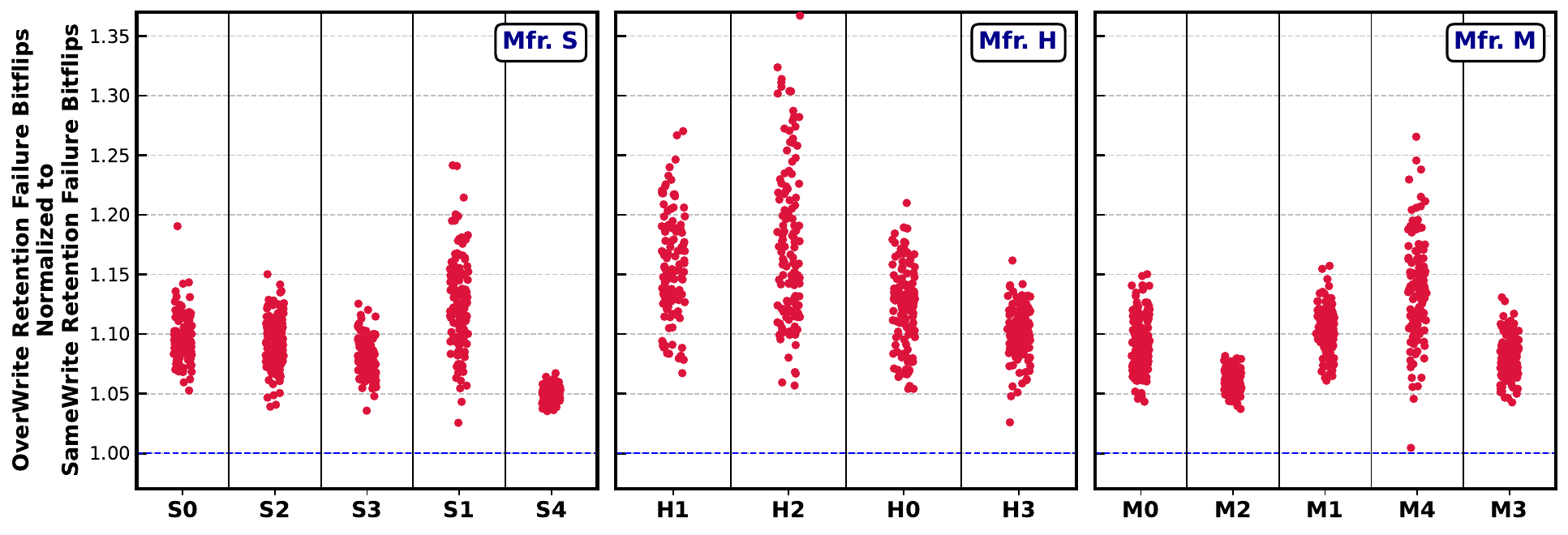}
    \caption{Distribution of the average number of DRAM retention failure bitflips from OverWrite normalized to SameWrite across all \arxivaccept{128 tested rows per module.}}
    \label{fig:ret_all}
\end{figure}

\figref{fig:ret_all} shows the distribution of the average number of DRAM retention failure bitflips from OverWrite normalized to SameWrite for all tested rows and modules. We highlight $y=1.0$ with the dashed blue line. We observe that when initializing \arxivaccept{a} tested row with the OverWrite pattern, we can induce more retention failure bitflips compared to using the SameWrite pattern for all tested DRAM \arxivaccept{rows and} chips across all three manufacturers. On average (geometric mean), OverWrite induces 10.4\% more bitflips compared to SameWrite (up to 36.7\% more).

\takeaway{DejaVu \arxivaccept{worsens} DRAM's vulnerability to retention failures.}

%% file: sections/05-sensitivity-to-tWR.tex
\section{Hypotheses on the Causes of DejaVu}
\arxivaccept{Based on our observations and prior works, we propose two non-mutually exclusive hypotheses on the potential physical mechanisms behind DejaVu.}
\label{sec:hypotheses}
\subsection{Charge Under-Restoration in the Capacitor}
\arxivaccept{Activating a DRAM row is \emph{destructive} for the information stored in the DRAM cell capacitor because the cell capacitance is much smaller than that of the bitline. After charge sharing, the cell capacitor voltage becomes very close to $V_{DD}/2$ instead of $V_{DD}$ or $0$. Therefore, after charge sharing, the DRAM row needs to stay open for the \arxivaccept{bitline sense amplifier (BLSA)} to \emph{restore} the charge level in the cell capacitors.} \arxivaccept{We hypothesize one of the reasons that causes DejaVu in DRAM is that overwriting the DRAM row with opposite data (OverWrite) may cause \arxivaccept{\emph{charge under-restoration}} in the DRAM cell capacitors compared to writing the same data twice (SameWrite).} 

\arxivaccept{We find two pieces of empirical evidence from our \arxivaccept{characterization results and testing methodology that potentially} support this hypothesis. First, Observation 4 shows that the reduction in \acmin is higher when we overwrite the victim row that was previously written with \texttt{0x00} with \texttt{0xFF} compared to overwriting the victim row that was previously written with \texttt{0xFF} with \texttt{0x00}. Since the DRAM access transistor is an NMOS, as the cell capacitor voltage increases (i.e., when writing \texttt{0xFF}), the Gate-Source ($V_{GS}$) voltage of the access transistor reduces, which reduces the current passing capability of the access transistor~\cite{Jang2017RefreshAware, luo2020clr} and makes \texttt{0xFF} more difficult to fully restore compared to \texttt{0x00}.} 

Second, the process of writing to all cache lines in the DRAM row is a serialized process. The cache lines that are written earlier have more effective charge restoration time compared to those written later in the DRAM row. For example, when writing to the entire DRAM row (i.e., 128 cache lines), the first cache block has $127\times$ tCCD\_L more effective write recovery time compared to the last cache block, \arxivaccept{potentially leading to a charge under-restoration in the later cache lines}. \arxivaccept{If we observe a significant concentration of the bitflips in the later cache lines of the victim DRAM row, then our observation that OverWrite worsens DRAM read disturbance compared to SameWrite can be explained as follows: the later cache lines in the victim DRAM row have much smaller effective charge restoration time with respect to the \emph{final} data pattern with SameWrite compared to OverWrite.} 

\arxivaccept{Unfortunately, due to limited observability and control at the DRAM chip level, it is difficult to directly identify the exact root cause(s) of DejaVu. Therefore, to gain chip-level insights into \arxivaccept{the \emph{charge under-restoration}} hypothesis, we empirically stress the hypothesis in \secref{sec:sen_tWR} and \secref{sec:spatial} by performing further characterizations that sweep \emph{additional} write-recovery time before sending \texttt{PRE} and study the spatial distribution of the \emph{initial} bitflips within the victim DRAM row, respectively.}

\subsection{Charge Trap State Changes in the Active Region}
Second, we hypothesize that the process of writing different \arxivaccept{data values} to the DRAM cell changes the charge trap states in the active regions where the DRAM access transistors are formed. These charge traps can trap electrons when the DRAM row is open, and then release them when the row is closed~\cite{yang2019trap, Jie2024Understanding, Zhou2024Unveiling, Zhou2024Understanding, Zhou2023Double}. \arxivaccept{Prior device-level studies on DRAM read disturbance~\cite{yang2019trap, Jie2024Understanding, Zhou2024Unveiling, Zhou2024Understanding, Zhou2023Double} identify charge trap assisted electron migration and injection into the victim cell as one of the major mechanisms behind RowHammer and RowPress.} 

\arxivaccept{We hypothesize that DejaVu may change the charge trap occupancy states in the active region that can affect \acmin in the following two ways. First, \arxivaccept{DejaVu-induced} changes in the charge trap occupancy states may change the effective threshold voltage \arxivaccept{ and the current passing capability} of the access transistors of the victim DRAM cells, affecting the subthreshold leakage current of the victim DRAM cells. }

\arxivaccept{Second, writing to the victim DRAM cells may also change the charge trap occupancy states for charge traps near the aggressor DRAM cells.} In modern $6F^2$ high density DRAM layout~\cite{schloesser20086f, yang2019trap, Zhou2023Double, nam2024dramscope}, two DRAM access transistors in two physically adjacent DRAM rows can share the same \emph{physical} active region. \arxivaccept{Therefore, charge traps whose \arxivaccept{occupancies are} perturbed by writing to the victim row can influence the read-disturbance-induced leakage caused by trap-assisted electron migration and injection~\cite{yang2019trap, Jie2024Understanding, Zhou2024Unveiling, Zhou2024Understanding, Zhou2023Double} of the physically adjacent aggressor row \emph{during} subsequent hammering.} 

\arxivaccept{Due to limited observability and control at the DRAM chip level, we characterize how DejaVu patterns affect the reliability of Processing-Using-DRAM (PUD) operations~\cite{seshadri2017ambit, hajinazar2021simdram, yuksel2024simultaneous, yuksel2024functionally} in~\secref{sec:pud} to provide empirical evidence for this \arxivaccept{\emph{charge trap state change}} hypothesis. PUD operations~\cite{seshadri2017ambit, hajinazar2021simdram, yuksel2024simultaneous, yuksel2024functionally} are highly sensitive to the current passing capability of the DRAM cell access transistors because PUD operations rely on a delicate charge sharing process involving the simultaneous activation of multiple input DRAM rows.}

\section{Deeper Characterization of DejaVu}
\label{sec:sensitivity}
\subsection{Sensitivity to Additional Write Recovery Time}
\label{sec:sen_tWR}
\begin{figure}[ht]
    \centering
    \includegraphics[width=0.9\linewidth]{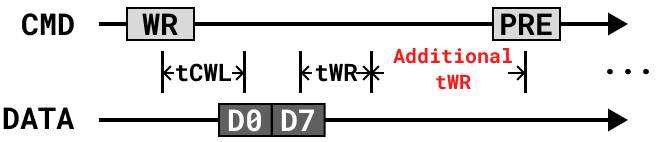}
    \caption{\arxivaccept{DejaVu with} additional Write Recovery Time.}
    \label{fig:tWR2_pattern}
\end{figure}

To investigate the \arxivaccept{potential impact of the difference in charge restoration \arxivaccept{(i.e., \emph{write recovery})} time on \acmin for DejaVu}, we modify the DejaVu pattern to include \emph{additional} wait times (i.e., additional write recovery time) for the \emph{second} write to the victim DRAM row. \arxivaccept{\figref{fig:tWR2_pattern} depicts the modified pattern.}

\begin{figure}[ht]
    \centering
    \includegraphics[width=\linewidth]{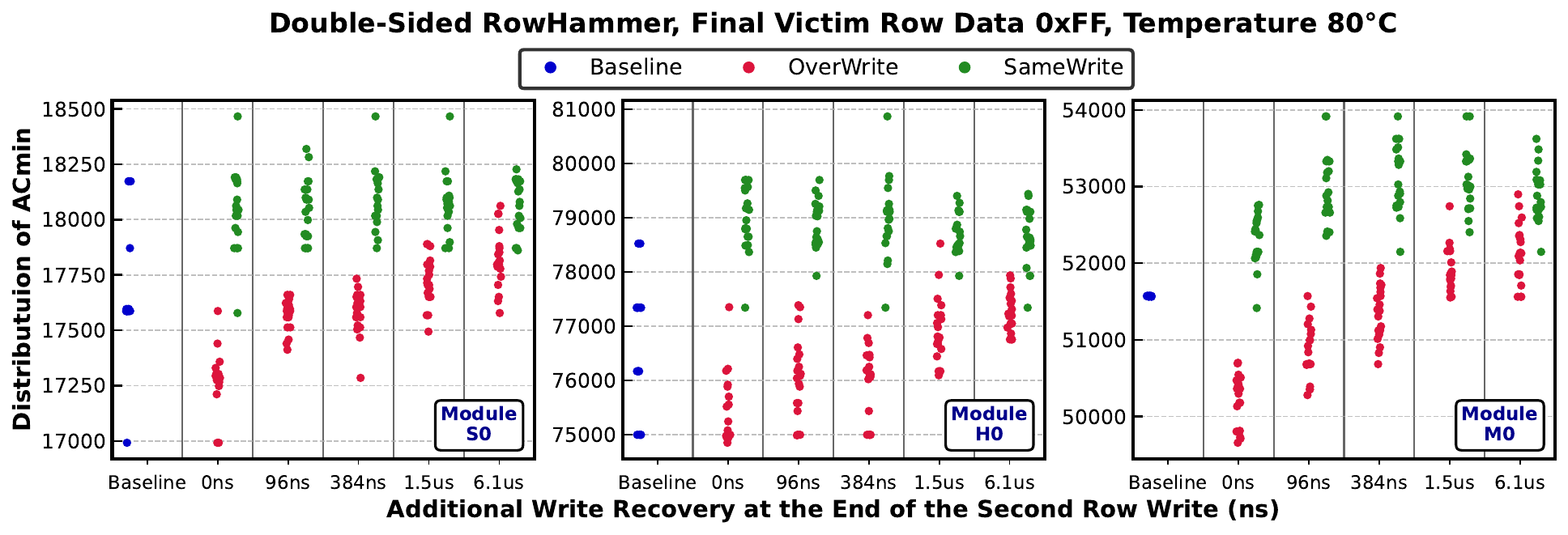}
    \caption{Distribution of the baseline and DejaVu Double-Sided RowHammer \acmin with different additional write recovery time to the second write; Victim data pattern \texttt{0xFF}, temperature $80^\circ\mathrm{C}$; Three representative victim rows \arxivaccept{from three representative modules from each of the three manufacturers}.}
    \label{fig:tWR2}
\end{figure}

\figref{fig:tWR2} shows the Double-Sided RowHammer \acmin distribution (y-axis) of the \arxivaccept{unmodified baseline pattern where we write to the victim row only once \arxivaccept{\emph{without}} any additional write recovery time} (blue) and the two DejaVu patterns (red and green) as the additional write recovery time (x-axis) to the second write of the DejaVu patterns increases. \arxivaccept{For the two DejaVu patterns, a write recovery time of 0ns corresponds to the unmodified DejaVu pattern (\figref{fig:wr_cmd_sequence}).} For clarity, we choose three representative victim rows, one each from a DRAM module from each of the three major manufacturers. 

\arxivaccept{We observe that as the additional write recovery time increases, the \acmin of the OverWrite pattern also increases, while the \acmin of the SameWrite pattern remains almost the same (S0 and H0), or slightly increases with a magnitude smaller than that of the OverWrite pattern (M0). Based on this observation, we believe the charge (under-)restoration hypothesis \emph{alone} is \emph{insufficient} to explain DejaVu. Although the increase in OverWrite's \acmin as additional write recovery time grows seems to agree with the charge restoration hypothesis, it is difficult to explain why \arxivaccept{there is only a very small change (if at all) in the \acmin of} SameWrite. Moreover, it is unlikely that it takes more than 6100ns (more than $400\times$ the standard tWR \arxivaccept{specified by the} JEDEC standard~\cite{jedec2017ddr4}) to fully restore the charge level in the victim cell (i.e., even after 6100ns, the \acmin of the OverWrite pattern is still lower than that of the SameWrite pattern).}

We also observe that the \acmin distribution of the baseline pattern shows a much larger spread compared to both DejaVu patterns. We hypothesize that since the baseline pattern writes to the victim row \arxivaccept{only once}, \arxivaccept{the actual physical process that occurs in the baseline pattern is much less consistent compared to the DejaVu patterns.}

\subsection{Spatial Distribution of RowHammer Bitflips Under DejaVu}
\label{sec:spatial}
\begin{figure}[ht]
    \centering
    \includegraphics[width=\linewidth]{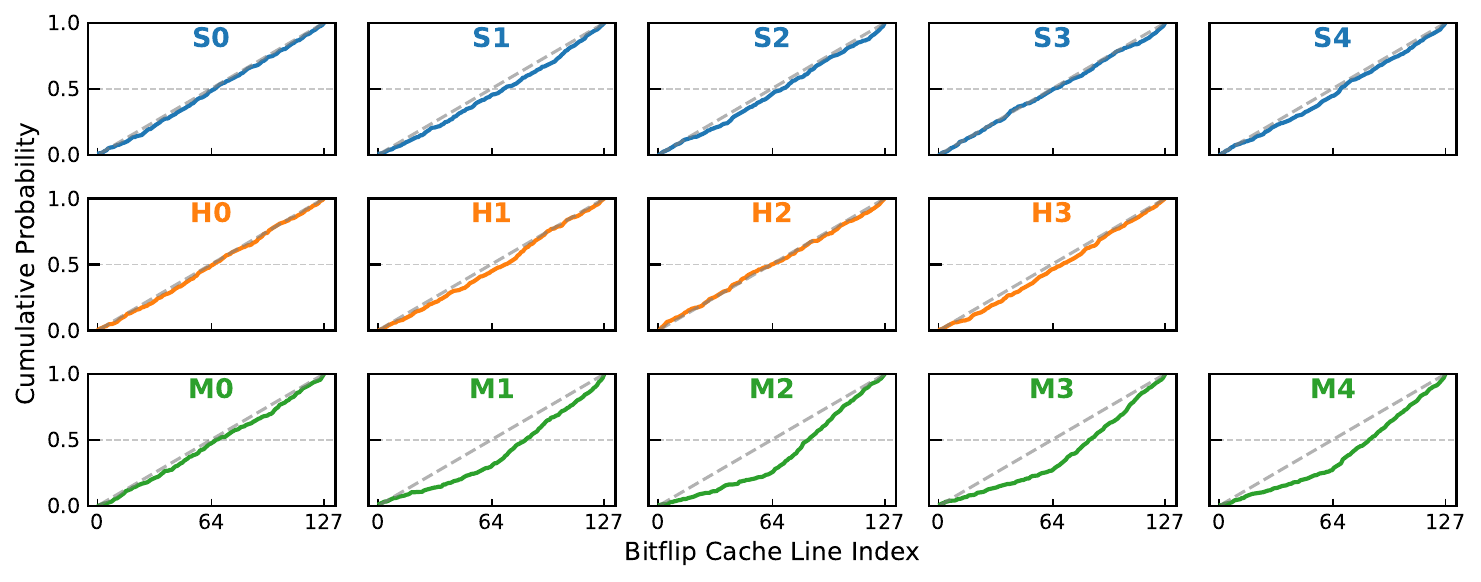}
    \caption{Cumulative probability distribution of all the initial Double-Sided RowHammer bitflips cache block indices for all tested victim rows, all chips, all data patterns, all temperatures.}
    \label{fig:CDF}
\end{figure}

To provide insights \arxivaccept{into} whether the imbalance of effective write recovery time of different cache lines in the row is the major factor behind DejaVu, we analyze the distribution of the indices of the cache lines where the \emph{initial} RowHammer bitflips appear in our tests.~\figref{fig:CDF} shows the cumulative probability of all the \emph{initial} Double-Sided RowHammer bitflips (i.e., those that flip at \acmin) cache block indices for all tested victim rows, all chips, all data patterns, and all temperatures. 

We observe that, for the majority of the modules tested (except for M1-M4\footnote{\arxivaccept{We attribute this to 1) potential half-row organization~\cite{nam2024dramscope} in the DRAM array, and 2) sophisticated inter- and intra-row interleaving of true- and anti-cells~\cite{khan2014efficacy, patel2017reaper, kraft2018improving, nam2024dramscope} that we could not fully reverse engineer.}}), there is no significant accumulation of the Double-Sided RowHammer bitflips in the cache lines towards the end of the victim row (i.e., the distribution follows the $y=x$ line). We hypothesize that this implies the charge under-restoration caused by the imbalanced effective write recovery time is not the \emph{only} mechanism involved in DejaVu.

\begin{figure}[ht]
    \centering
    \includegraphics[width=\linewidth]{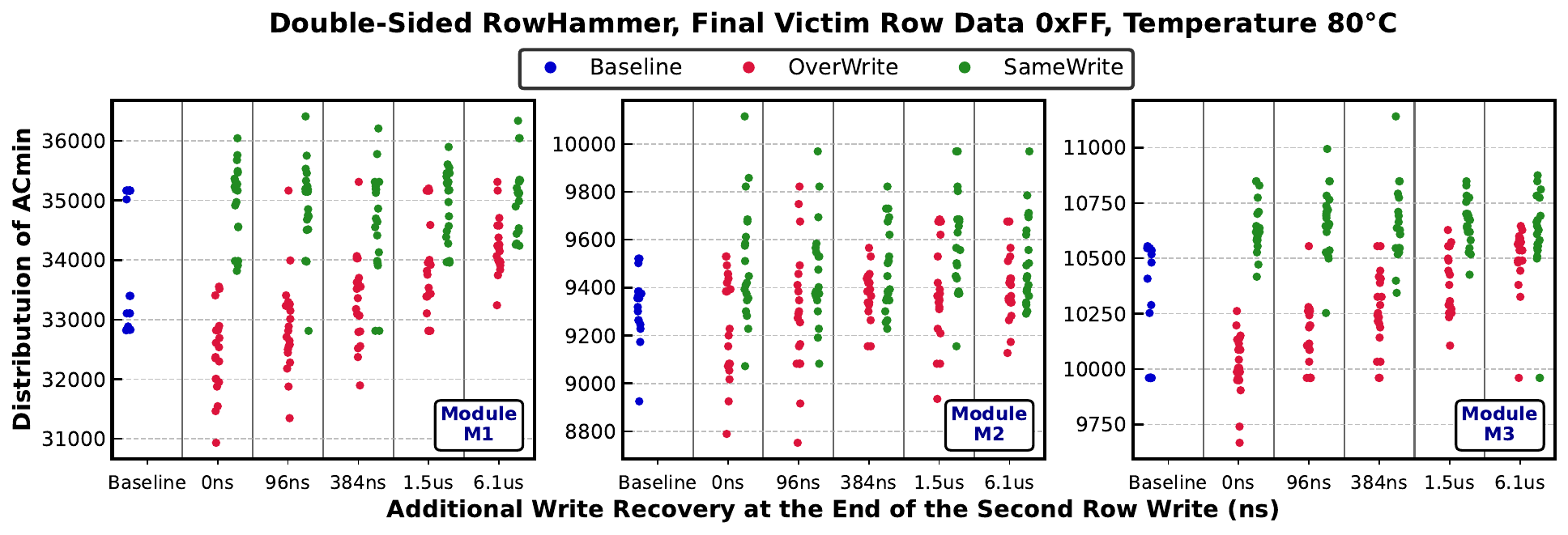}
    \caption{Distribution of the baseline and DejaVu Double-Sided RowHammer \acmin with different additional write recovery time to the second write; Victim data pattern \texttt{0xFF}, temperature $80^\circ\mathrm{C}$; Three representative victim rows from Modules M1, M2, and M3.}
    \label{fig:tWR2_M123}
\end{figure}

\figref{fig:tWR2_M123} \arxivaccept{shows the same Double-Sided RowHammer \acmin distribution of the OverWrite and SameWrite patterns compared to the baseline pattern} as \figref{fig:tWR2} but for three modules that have the Double-Sided RowHammer bitflip cache block indices more concentrated in the second half of the victim row \arxivaccept{(i.e., the distribution is below the $y=x$ line)}. Interestingly, \arxivaccept{these modules} also behave the same as the other modules in the additional write recovery characterization in~\secref{sec:sen_tWR}. We call for more detailed study to understand and explain this observation.

\subsection{Data Patterns}
We test the \texttt{0x00} and \texttt{0xFF} data patterns in the previous sections because there exists no solid reverse-engineering methodology to uncover the mapping between the DQ pins and the bitlines in the DRAM cell array. Without such a methodology, there is no guarantee that a checkerboard pattern (i.e., \texttt{0xAA} and \texttt{0x55}) sent to the DRAM will end up being the same checkerboard pattern in the DRAM cell array. Therefore, to avoid introducing uncontrolled factors in our results, we choose to use the \texttt{0xFF} and \texttt{0x00} patterns. In this section, we present data for selected experiments with the checkerboard \texttt{0xAA} and \texttt{0x55} data patterns and show that our major observations and takeaways still hold for the checkerboard data patterns.

\begin{figure}[h]
    \centering
    \includegraphics[width=\linewidth]{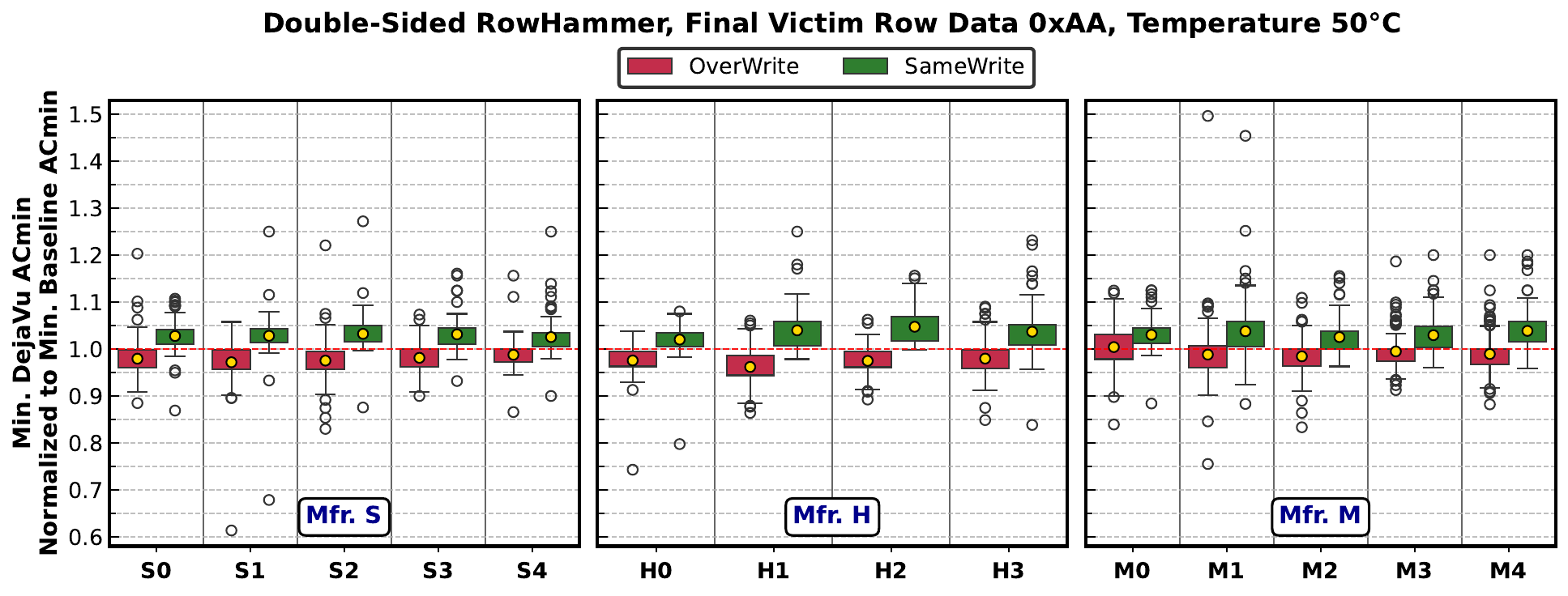}
    \caption{Minimum DejaVu \acmin normalized to minimum baseline \acmin, Double-Sided RowHammer, victim data \texttt{0xAA}, $50^\circ\mathrm{C}$. \arxivaccept{Distribution across all 128 tested rows per module.}}
    \label{fig:RH_acmin_min_allrows_AA}
\end{figure}

\begin{figure}[h!]
    \centering
    \includegraphics[width=\linewidth]{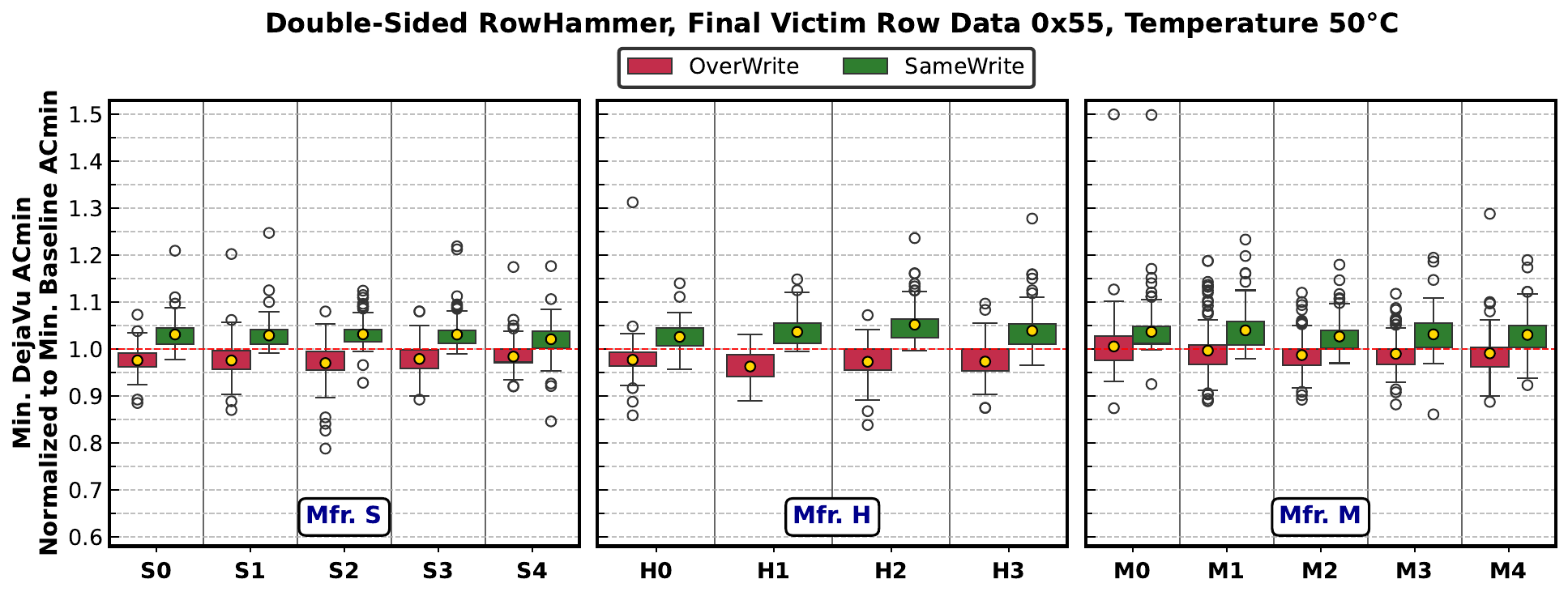}
    \caption{Minimum DejaVu \acmin normalized to minimum baseline \acmin, Double-Sided RowHammer, victim data \texttt{0x55}, $50^\circ\mathrm{C}$. \arxivaccept{Distribution across all 128 tested rows per module.}}
    \label{fig:RH_acmin_min_allrows_55}
\end{figure}

~\figsref{fig:RH_acmin_min_allrows_AA} and~\ref{fig:RH_acmin_min_allrows_55} show the distribution of the \emph{minimum} DejaVu \acmin normalized to the \emph{minimum} baseline \acmin for victim row data patterns \texttt{0xAA} and \texttt{0x55}, respectively, using the same methodology as in~\figsref{fig:RH_acmin_min_allrows_00} and~\ref{fig:RH_acmin_min_allrows_FF}. \arxivaccept{We find that OverWrite reduces the minimum \acmin by 2.1\% (2.3\%), 2.7\% (2.9\%), and 0.8\% (0.6\%) compared to the baseline where the victim row is written only once for Mfr. S, H, and M, respectively, for the \texttt{0xAA} (\texttt{0x55}) victim data pattern.} We conclude that our major observations on how DejaVu changes \acmin compared to the baseline case where the victim row is written only once (Observations 1-3 in~\secref{sec:rh_acmin_results}) still hold for the checkerboard data patterns.

\begin{figure}[!t]
    \centering
    \includegraphics[width=\linewidth]{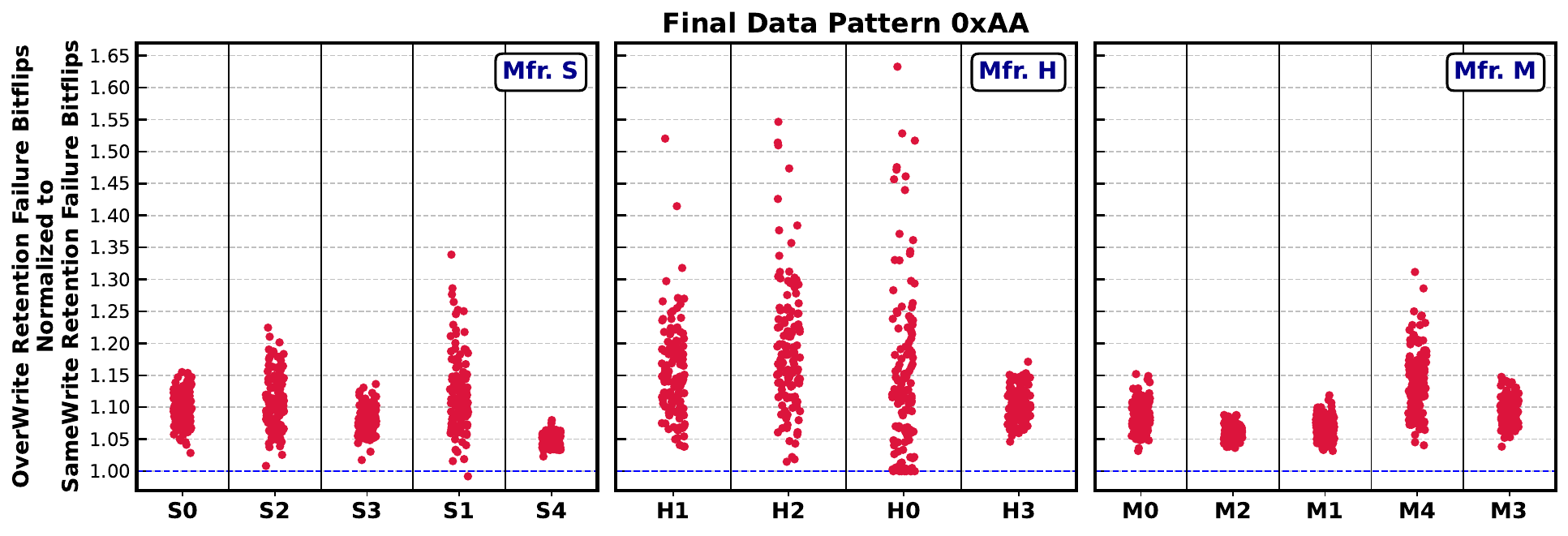}
    \caption{Example distribution of the average number of DRAM retention failure bitflips from OverWrite normalized to SameWrite (\texttt{0xAA} data pattern) \arxivaccept{across all 128 tested rows per module.}}
    \label{fig:ret_all_AA}
\end{figure}

\begin{figure}[!t]
    \centering
    \includegraphics[width=\linewidth]{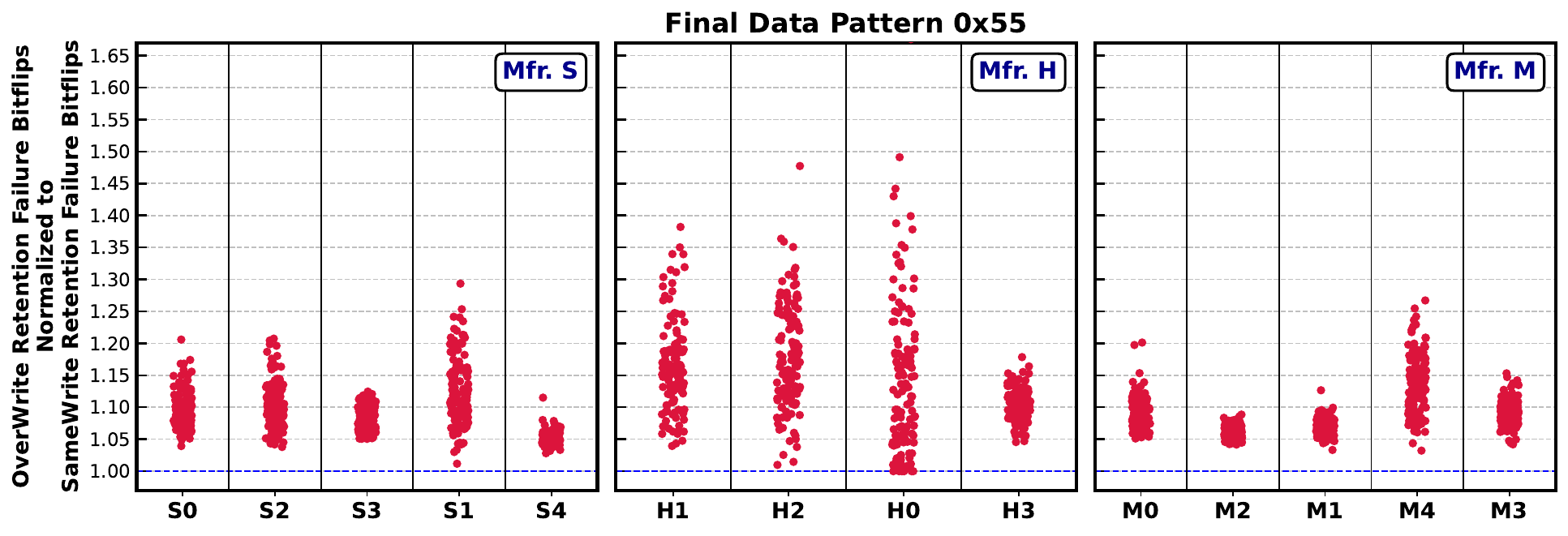}
    \caption{Example distribution of the average number of DRAM retention failure bitflips from OverWrite normalized to SameWrite (\texttt{0x55} data pattern) \arxivaccept{across all 128 tested rows per module.}}
    \label{fig:ret_all_55}
\end{figure}

~\figsref{fig:ret_all_AA} and~\ref{fig:ret_all_55} show the distribution of the average number of DRAM retention failure bitflips from OverWrite normalized to SameWrite for all tested rows and modules for data patterns \texttt{0xAA} and \texttt{0x55}, respectively, using the same methodology as in~\figref{fig:ret_all}. \arxivaccept{We observe that on average, OverWrite increases the number of retention failure bitflips by 10.7\% and 10.6\% (up to 123.2\% and 67.6\%) for the \texttt{0xAA} and \texttt{0x55} data patterns, respectively, compared to SameWrite.} We conclude that our major takeaway on how DejaVu enhances DRAM's vulnerability to retention failures (Takeaway 4) still holds for the checkerboard data patterns.

\begin{figure}[!t]
    \centering
    \includegraphics[width=\linewidth]{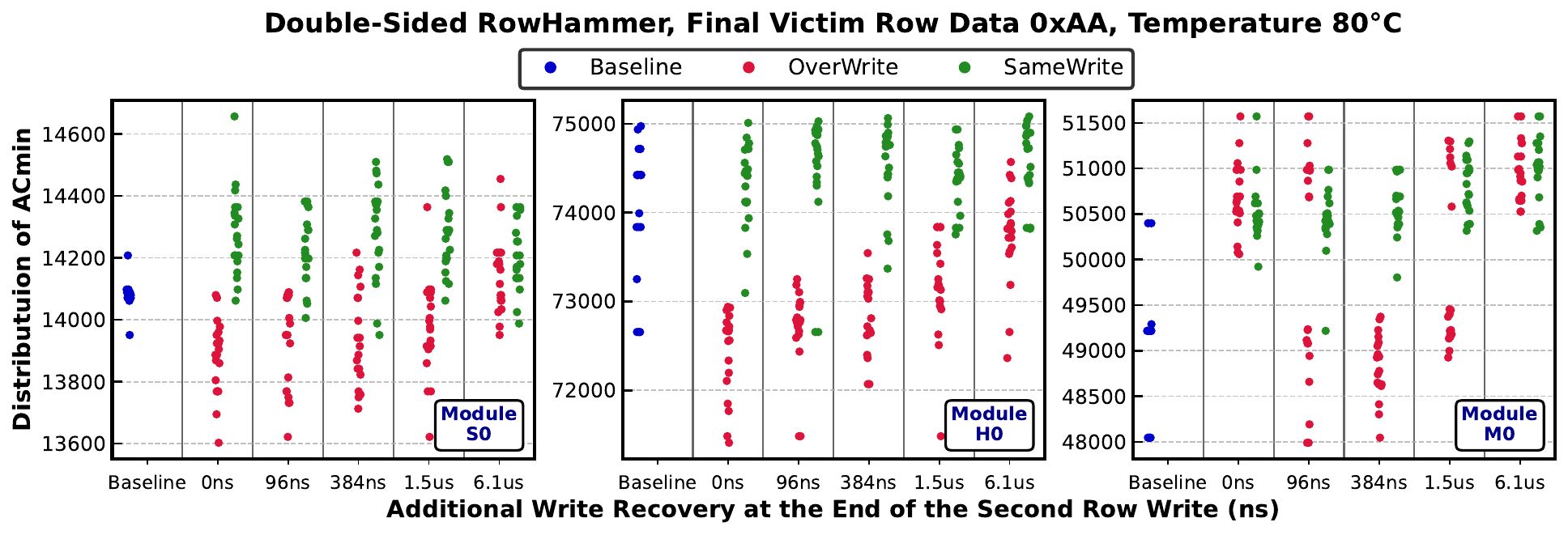}
    \caption{Distribution of the baseline and DejaVu Double-Sided RowHammer \acmin with different additional write recovery time to the second write; Victim data pattern \texttt{0xAA}, temperature $80^\circ\mathrm{C}$; Three representative victim rows \arxivaccept{from three representative modules from each of the three manufacturers}.}
    \label{fig:tWR_AA}
\end{figure}

\begin{figure}[!t]
    \centering
    \includegraphics[width=\linewidth]{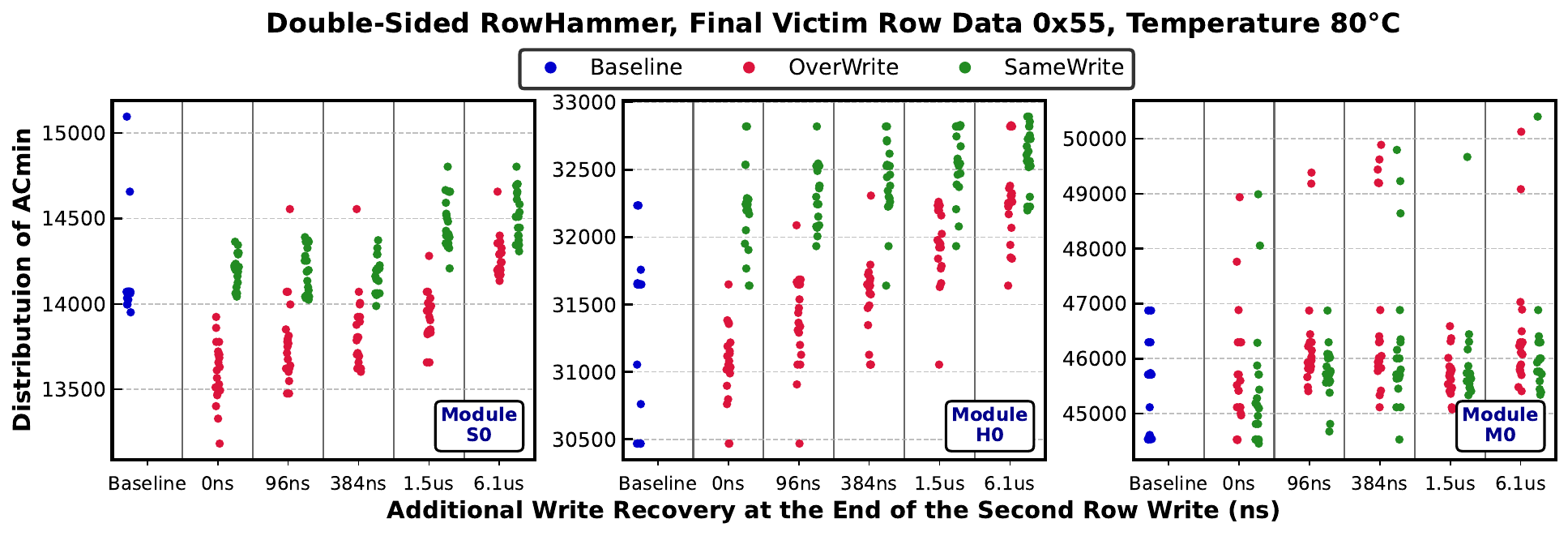}
    \caption{Distribution of the baseline and DejaVu Double-Sided RowHammer \acmin with different additional write recovery time to the second write; Victim data pattern \texttt{0x55}, temperature $80^\circ\mathrm{C}$; Three representative victim rows \arxivaccept{from three representative modules from each of the three manufacturers}.}
    \label{fig:tWR_55}
\end{figure}

{\looseness=-1
~\figsref{fig:tWR_AA} and~\ref{fig:tWR_55} show the Double-Sided RowHammer \acmin distribution (y-axis) of the baseline pattern (blue) and the two DejaVu patterns (red and green) as the additional write recovery time (x-axis) to the second write of the DejaVu patterns increases for data patterns \texttt{0xAA} and \texttt{0x55}, respectively, using the same methodology as in~\figref{fig:tWR2}. We reach the same conclusion as we do in~\secref{sec:sen_tWR} that charge \arxivaccept{under-restoration} is likely \emph{not} the major underlying mechanism \arxivaccept{that fully explains DejaVu}.\par}

%% file: sections/06-PUD.tex
\section{Impact of DejaVu on PUD Reliability}
\label{sec:pud}
To further investigate the hypothesis that the DejaVu patterns (i.e., OverWrite and SameWrite) affect the charge trap states in the silicon substrates of the DRAM cell access transistors, we experimentally examine and characterize how DejaVu changes the reliability and stability of Processing-Using-DRAM (PUD)~\cite{seshadri2015fast, seshadri2017ambit, seshadri2018rowclone, hajinazar2021simdram, deoliveira2024mimdram, yuksel2024simultaneous, mutlu2024memory, mutlu2025memory, oliveira2025proteus, Yuksel2025PuDHammer, yuksel2024functionally, gao2019computedram, gao2022frac, olgun2021quactrng, bostanci2022drstrange, tokuda2026clutch} operations \arxivaccept{on real COTS DRAM chips~\cite{yuksel2024simultaneous, olgun2021pidram, Yuksel2025PuDHammer, yuksel2024functionally, gao2019computedram, gao2022frac, olgun2021quactrng, tokuda2026clutch}}. \arxivaccept{Recently demonstrated} PUD operations \arxivaccept{on real commercial-off-the-shelf (COTS) DRAM chips} rely on a delicate charge sharing process involving \arxivaccept{the simultaneous activation of} multiple input DRAM rows \arxivaccept{in the same subarray~\cite{yuksel2024simultaneous,yuksel2024functionally}}, which is highly sensitive to the current passing capability of the DRAM cell access transistors that is affected by the charge trap states in the silicon substrate of the transistors.
\subsection{Testing Methodology}

\noindent\textbf{Finding Subarray Boundaries.}
To understand \arxivaccept{the effect of DejaVu on} the computational capability of PUD operations performed in a DRAM subarray, we follow the methodology used in prior works to \arxivaccept{first} identify subarray boundaries~\cite{yaglikci2022hira, gao2019computedram, gao2022frac, olgun2021quactrng, olgun2021pidram,olgun2023dram, yuksel2024functionally, yuksel2024simultaneous}. We leverage the observation that it is possible to copy a row's data to another row (i.e., RowClone operation~\cite{seshadri2013rowclone,olgun2021pidram}) within the same subarray by leveraging the shared bitlines. We repeatedly perform RowClone across all tested rows. If we can copy a row's data to another row, we conclude that the source row and the destination row are in the same subarray. Based on this observation, we reverse engineer the subarray boundaries and determine which rows are in the same subarray.

\noindent\textbf{Finding Simultaneously Activated Rows.} 
Prior works~\cite{olgun2021quactrng, yuksel2024simultaneous, yuksel2024functionally} show that issuing an \texttt{ACT}-\texttt{PRE}-\texttt{ACT} command sequence with violated timings and \arxivaccept{followed by} a \texttt{WR} command overwrites the simultaneously activated rows with data supplied with the \texttt{WR} command. We follow the same methodology and reverse engineer the simultaneously activated rows with the \texttt{ACT}-\texttt{PRE}-\texttt{ACT} command sequence for every row address in a tested subarray. Similar to prior works~\cite{yuksel2024simultaneous, yuksel2024functionally}, we observe that real DRAM chips can activate 2, 4, 8, 16, and 32 rows in the same subarray.

\noindent\textbf{Bitwise Majority-of-Three Experiments.}
We follow the same methodology of prior work that leverages simultaneous multiple-row activation to perform in-DRAM majority-of-three (MAJ3) experiments~\cite{yuksel2024simultaneous}. 
We perform MAJ3 operations using 16- and 32-row activation (i.e., simultaneously activating 16 and 32 rows), as DRAM performs MAJ3 using 16- and 32-row activation much more reliably than using 4- and 8-row activation~\cite{yuksel2024simultaneous}. {\arxivaccept{We use random data patterns} (i.e., each input value of MAJ3 is generated randomly)} \arxivaccept{that both 1) \arxivaccept{likely} match real program input data better than all \texttt{0x00}s and all \texttt{0xFF}s, and 2) maximize inter-bitline interference during simultaneous multi-row activation}.

To perform MAJ3 operation with N-row activation (i.e., simultaneously activating N rows, where N can be either 16 or 32), we replicate each MAJ3 input operand (i.e., a total of three input operands) $\lfloor N/3 \rfloor$ times. For example, if we perform MAJ3 with 32-row activation, we replicate each input ten (i.e., $\lfloor 32/3 \rfloor = 10$) times. Since the number of DRAM rows opened is not a multiple of 3, we initialize the remaining activated DRAM rows in a way that they do not contribute to bitline voltage, \arxivaccept{using the Frac operation~\cite{gao2022frac}.} For example, if we perform MAJ3 with 32-row activation, we perform the Frac operation on the remaining two DRAM rows.

\noindent\textbf{Number of Tested Instances.} To maintain a reasonable testing time, we randomly select a total of three subarrays in one bank per DRAM module. Within each subarray, we randomly test 100 different groups of rows that are simultaneously activated, each for 16- and 32-row activation.

\noindent\textbf{Reliability Metric.} \arxivaccept{We define \emph{bitline failure rate} as a metric to evaluate the reliability of MAJ3 operations.} The bitline failure rate refers to the percentage of DRAM cells (bitlines) that produce {an \emph{incorrect}} output in {at least one of the} trials {(we perform a total of 1K trials)}. Even if a bitline produces an incorrect result \arxivaccept{\emph{just once}}, we refer to this {DRAM} bitline as an \emph{unstable bitline} that \emph{cannot} be used to perform MAJ3 operations. For example, if a MAJ3 operation has a 25\% bitline failure rate, it means that 75\% of the {DRAM} bitlines always produce correct results in the simultaneously activated rows and can be used to reliably perform that operation.

\subsection{Results}
\begin{figure}[ht]
    \centering
    \includegraphics[width=\linewidth]{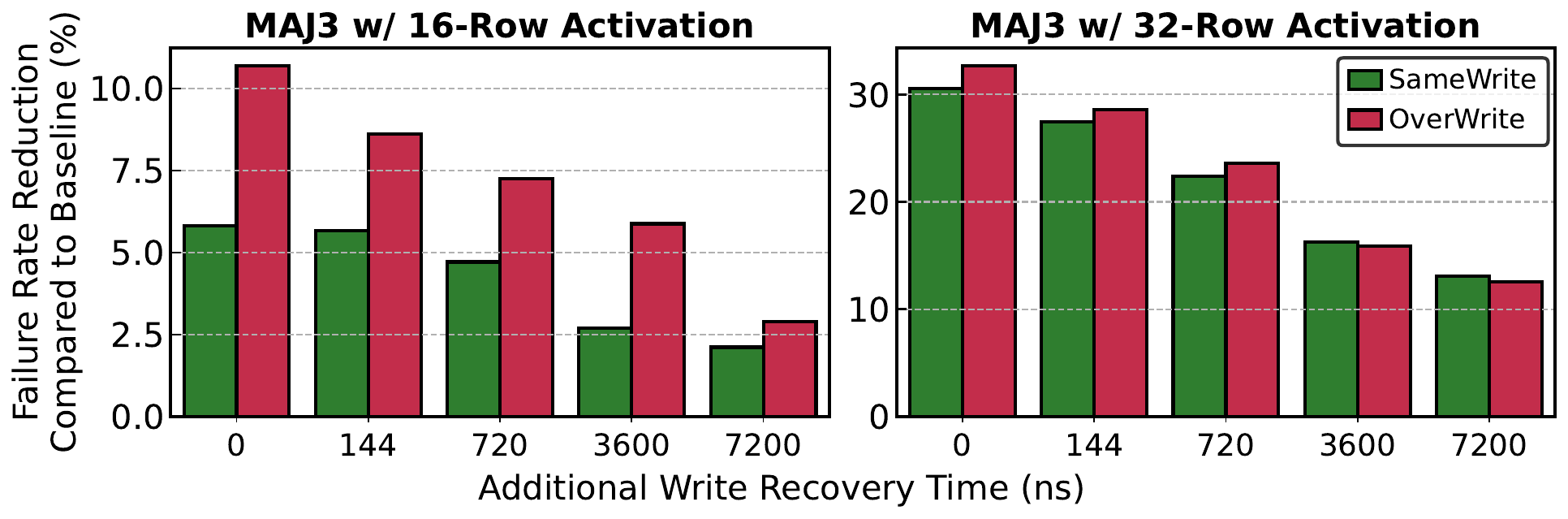}
    \caption{Bitline failure reduction leveraging DejaVu compared to baseline.}
    \label{fig:pud_memory_effect}
\end{figure}

\figref{fig:pud_memory_effect} shows the reduction in bitline failure rate (y-axis) as we leverage DejaVu (SameWrite and OverWrite) to initialize the \arxivaccept{16 or 32 (left and right plots, respectively)} DRAM rows involved in \arxivaccept{the} MAJ3 operation from all tested row groups in all tested DRAM chips as we change the \emph{additional} write recovery time (x-axis) we set between the arrival of the data burst of the last cache block write to the row and the next \texttt{PRE} command \arxivaccept{(as in~\figref{fig:tWR2_pattern})}.

We make the following two observations. First, when the additional write recovery time is 0, OverWrite reduces the bitline failure rate by 10.7\% (MAJ3 - 16 row) and 32.7\% (MAJ3 - 32 row) compared to the baseline. SameWrite reduces the bitline failure rate by 5.8\% (MAJ3 - 16 row) and 30.6\% (MAJ3 - 32 row) compared to the baseline. \arxivaccept{\emph{As such, using the DejaVu effect can make PUD operations much more reliable.}} Second, as the additional write recovery time increases, the bitline failure rate reduction caused by DejaVu decreases.

\subsection{Hypotheses}
Due to limited observability from the DRAM chip's level, we provide two hypotheses to explain 1) why both DejaVu access patterns (i.e., SameWrite and OverWrite) improve the reliability of PUD operations compared to the baseline case where the DRAM row(s) are written to \arxivaccept{only once}, and 2) why the OverWrite pattern increases the reliability of PUD operations more than that of SameWrite.

First, we hypothesize that both DejaVu access patterns make the charge trap occupancy states in the silicon substrate of the DRAM cell access transistors more uniform compared to the baseline case. In the baseline case, since the DRAM row(s) are written to \arxivaccept{only once}, the charge trap occupancy states are more random compared to DejaVu. A more uniform distribution of charge trap occupancy states reduces the variations in the current passing capabilities of the DRAM access transistors, \arxivaccept{which improves the reliability of PUD operations.}

Second, we hypothesize that OverWrite \arxivaccept{\emph{increases}} the current passing capabilities of the DRAM cell access transistors, making the charge sharing process of PUD operation more robust. \arxivaccept{This hypothesis empirically agrees with our Observation 2 that OverWrite consistently reduces \acmin (i.e., the leakage current caused by RowHammer is higher when the victim row is initialized with OverWrite).}

From our results and \arxivaccept{hypotheses}, we call for more research, especially \arxivaccept{research to study} the device-level operation and physical mechanisms of DRAM, to 1) fundamentally understand the root causes of DejaVu \arxivaccept{and its effects on both read disturbance and PUD operations}, and 2) reduce the variation of the current passing capabilities of the DRAM access transistors to enable fundamentally more reliable PUD operations.

%% file: sections/08-Implications.tex
\section{Implications of DejaVu on DRAM Testing and Characterization Methodology}
\subsection{Finding the Worst-Case Read Disturbance Threshold}
Our extensive characterization results from~\secref{sec:rh_acmin_results} to~\secref{sec:sensitivity} demonstrate how DejaVu (especially the OverWrite pattern) can exacerbate DRAM read disturbance and retention failure bitflips. Since data update is a fundamental operation and very common in real system operations, DRAM testing and characterization methodologies should account for the DejaVu effects to enable safer and more robust DRAM operation. To this end, we propose incorporating the following practices into DRAM testing and characterization methodologies for read disturbance and retention failure bitflips:
\begin{enumerate}
    \item Initialize the victim data using OverWrite (i.e., first write opposite data, then write the actual test data) to capture the DejaVu-induced reduction of \acmin and the increase of retention failure bitflips.
    \item Test at individual \arxivaccept{cache block (i.e., column)} granularity to minimize \arxivaccept{potential differences} in effective charge restoration (write recovery) time when testing at a whole-row granularity.
    \item Repeat each measurement multiple times to increase the coverage of outliers.
\end{enumerate}

\subsection{Exploring the Effect of Data Patterns on Read Disturbance}
When the goal is to explore how different data patterns affect DRAM read disturbance, it is important to \arxivaccept{avoid any incorrect conclusion} as a result of accidentally inducing DejaVu-caused difference in \acmin. Listing~\ref{lst:example} demonstrates a typical scenario where DejaVu is \emph{accidentally} induced when initializing a continuous range of DRAM rows using a for loop.

\begin{figure}[tbp]
\centering
\begin{minipage}{0.9\columnwidth}
\begin{lstlisting}[style=pythoncompact,basicstyle=\ttfamily\footnotesize,caption={Pseudocode of a DRAM testing program that accidentally induces DejaVu \arxivaccept{effects}.},label={lst:example}]
def init_row_range(R_start, R_end, data):
    # Initialize a continuous range of rows
    for i in range(R_start, R_end+1):
      write_row(i, aggr_data)

##################################################
# Test data pattern dependency of R and R-2, R+2 #
##################################################
victim = R
victim_data = 0x00
aggressor_radius = 2

# Case 1: R-2 and R+2 initialized to store
# same as victim data
init_row_range(R-2, R+2, victim_data)
write_row(R, victim_data) # Accidental SameWrite
doublesided_hammer(ac, R-1, R+1)
check_bitflips(victim_data)

# Case 2: R-2 and R+2 initialized to store
# inverse of victim data
init_row_range(R-2, R+2, ~victim_data)
write_row(R, victim_data) # Accidental OverWrite
doublesided_hammer(ac, R-1, R+1)
check_bitflips(victim_data)
\end{lstlisting}
\end{minipage}
\end{figure}

The goal of Listing~\ref{lst:example} is to test, given a victim row R, whether the data pattern at rows R-2 and R+2 affects the \acmin of Double-Sided RowHammer (with aggressor rows R-1 and R+1) where prior works show that the data pattern at R-1 and R+1 matters. Since all 5 DRAM rows involved in the tested pattern are continuous, and the 4 non-victim rows all have the same data pattern, it is natural to use a range-based loop to initialize all the rows \arxivaccept{(including both the victim row and the 4 non-victim rows)} first (line 14, 20), and then initialize the single victim row afterwards (line 15, 21). However, doing so accidentally induces SameWrite (line 15) and OverWrite (line 21) on the victim row. As a result, Test Case 2 will give a lower \acmin than Test Case 1, and testers \arxivaccept{might mistakenly} interpret this difference in \acmin as evidence of data pattern dependency at rows R-2 and R+2.

To avoid \arxivaccept{such unintended effects or incorrect conclusions due to the existence} of DejaVu, when exploring how different data patterns affect DRAM read disturbance, \arxivaccept{we recommend that} future DRAM testing methodologies should \arxivaccept{avoid overwriting} the victim row. Instead, the victim row should always be initialized by writing the same data to it twice \arxivaccept{(i.e., using the SameWrite pattern)}.

%% file: sections/08a-mitigation.tex
\section{Impact on Read Disturbance Mitigations}
\label{sec:mitigation}
Existing read disturbance mitigations~\arxivaccept{\mitigatingRowHammerAllCitations} often rely on a configured read disturbance threshold ($N_{RH}$) to determine potential aggressor rows and prevent read disturbance bitflips by performing mitigative actions (i.e., refreshing potential victim rows). Our empirical results and analyses show that DejaVu can cause a difference in the observed read disturbance threshold. This can result in misconfigurations in the existing read disturbance mitigation techniques. We evaluate two read disturbance mitigation techniques \arxivaccept{(PARA~\cite{kim2014flipping} and PRAC~\cite{jedec2024ddr5, canpolat2024understanding, canpolat2025chronus})} to show the impact of DejaVu-caused differences in system performance with these mitigation techniques. \arxivaccept{We implement and evaluate PARA and PRAC in Ramulator 2.0~\cite{kim2016ramulator, luo2023ramulator2,ramulator2github, bostanci2026cleaning}. {We use 57 single-core workloads from SPEC CPU2006~\cite{spec2006}, SPEC CPU2017~\cite{spec2017}, TPC~\cite{tpcweb}, MediaBench~\cite{fritts2009media}, and YCSB~\cite{ycsb} to evaluate 60 random four-core workload mixes.}}

\begin{figure}[ht]
    \centering
    \includegraphics[width=\linewidth]{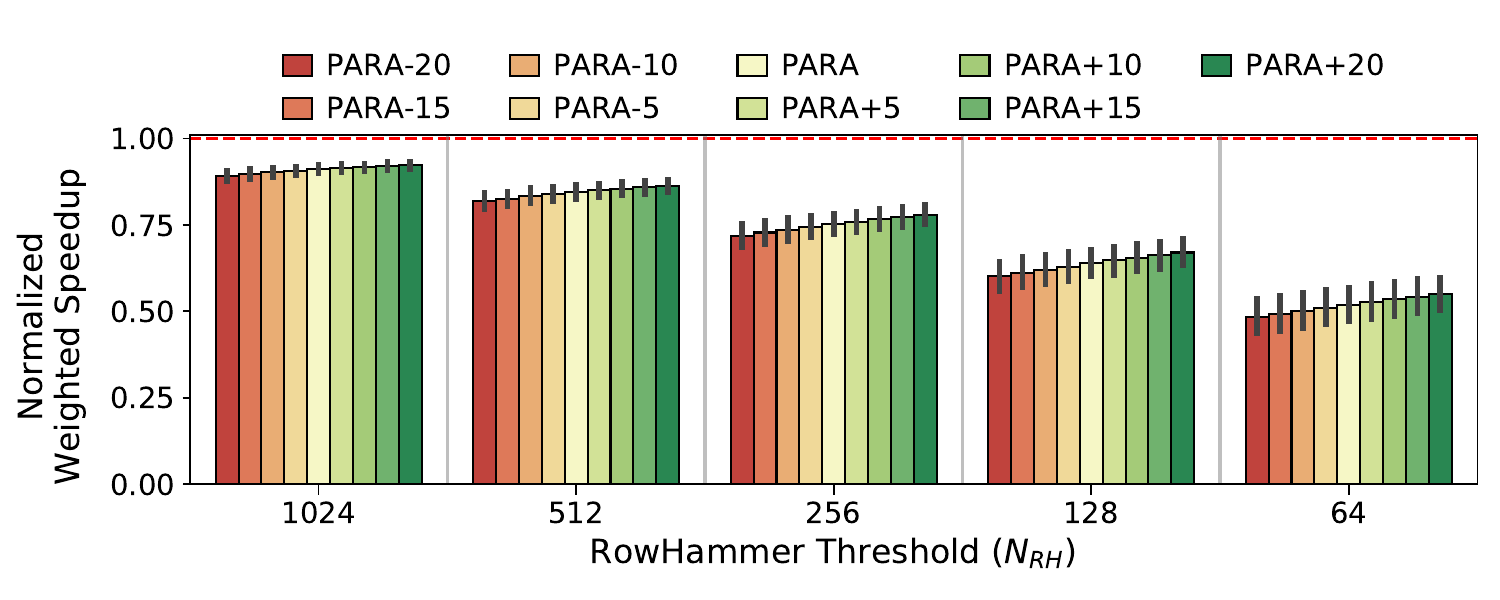}
    \caption{System performance with PARA with DejaVu caused $N_{RH}$ difference normalized to a baseline with no read disturbance mitigation technique.}
    \label{fig:para}
\end{figure}

\begin{figure}[ht]
    \centering
    \includegraphics[width=\linewidth]{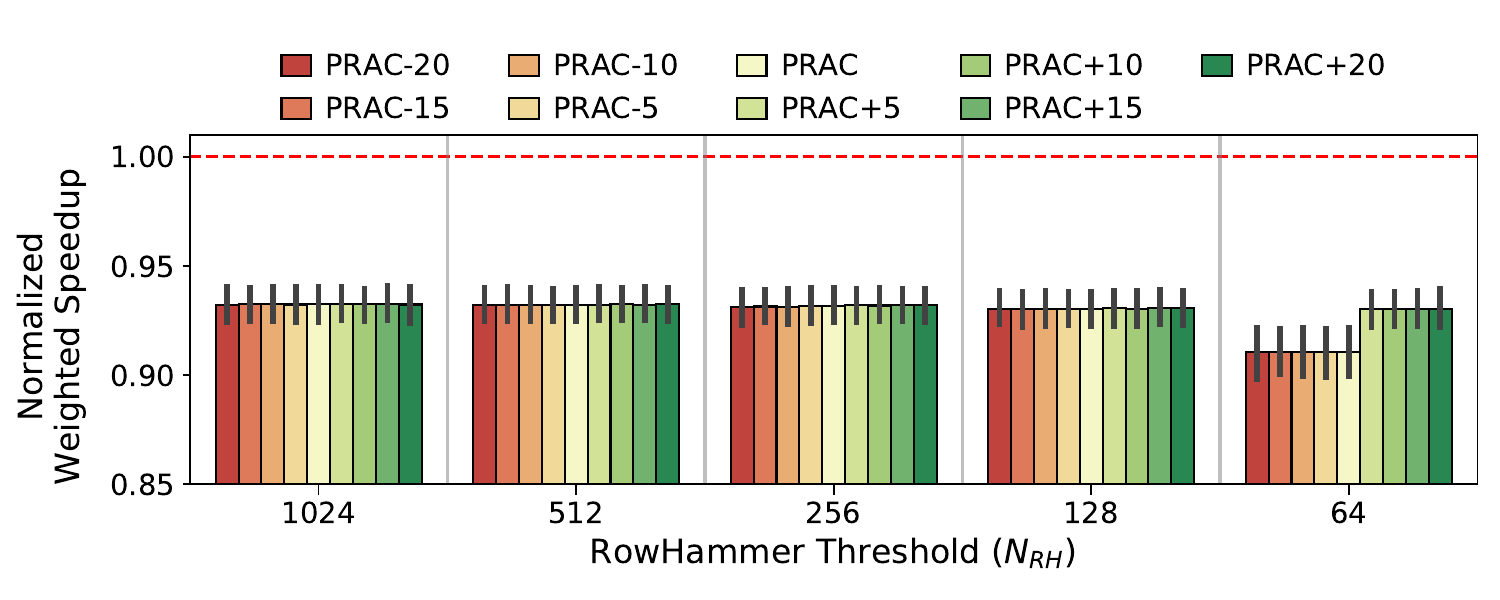}
    \caption{System performance with PRAC with DejaVu caused $N_{RH}$ difference normalized to a baseline with no read disturbance mitigation technique.}
    \label{fig:prac}
\end{figure}

\figsref{fig:para} and~\ref{fig:prac} show system performance {with} {PARA and PRAC}, respectively, with potential DejaVu-caused differences in $N_{RH}$ configuration, normalized to the baseline system that does
\emph{not} implement read disturbance mitigation, across 60 four-core workload mixes.  PARA~\cite{kim2014flipping} prevents read disturbance bitflips by
determining the target row of a DRAM activate command as an aggressor row
{based on a probability} {that is determined based on the
configured $N_{RH}$ value} and preventively refreshing {the aggressor row's}
neighbors. PRAC~\cite{jedec2024ddr5, canpolat2024understanding, canpolat2025chronus} tracks the
activation count of an aggressor row using \arxivaccept{in-DRAM per-row counters} and preventively
refreshes {the aggressor row's} neighbors before the activation count
reaches the {configured} read disturbance threshold. We sweep the potential increase and decrease in the $N_{RH}$ configuration from 20\% reduction to 20\% increase. We annotate each configuration with a DejaVu-caused difference in the form of \textit{\{Mitigation\}}\textit{\{Difference\}} where \emph{Difference} is the percentage of the $N_{RH}$ increase or reduction. The x-axis shows five different read disturbance threshold values ($N_{RH}$). 

We make two key observations. First, DejaVu-caused reduction in $N_{RH}$ results in increased performance overheads for both mitigation techniques across all tested $N_{RH}$ values. For example, at $N_{RH}$=64, 20\% reduction degrades the system performance with PARA by 6.3\% on average across all tested workloads. The performance overhead caused by DejaVu-caused difference is smaller for PRAC compared to PARA for the tested $N_{RH}$ values. This is because PRAC's performance overhead is mainly caused by the increased DRAM timing parameters for these $N_{RH}$ values~\cite{canpolat2024understanding,canpolat2025chronus}. 

Second, DejaVu-caused increase in $N_{RH}$ results in improved system performance with both mitigation techniques across all tested $N_{RH}$ values. For example, at $N_{RH}$=64, 20\% increase improves the system performance with PARA by 7.8\% and with PRAC by 2.1\%, on average across all tested workloads. 

We conclude that DejaVu-caused differences in $N_{RH}$ can result in significant system performance difference \arxivaccept{when read disturbance mitigation techniques are properly configured to account for DejaVu  effects.}

%% file: sections/08c-related_works.tex
\section{Related Work}
To our knowledge, this is the first work to experimentally demonstrate and characterize how the data \emph{previously} written to DRAM cells affects DRAM's vulnerability to read disturbance and the reliability of Processing-Using-DRAM (PUD) operations. In this section, we discuss other works on experimental characterization and demonstration of DRAM read disturbance and PUD capabilities.

\noindent\textbf{Read Disturbance Characterization.} Many works~\cite{orosa2021deeper,kim2020revisiting,kim2014flipping, yaglikci2022understanding, lim2017active, park2016statistical, park2016experiments, ryu2017overcoming, yun2018study, lim2018study, luo2023rowpress, lang2023blaster, yaglikci2024svard, nam2024dramscope, olgun2023hbm, olgun2025variable, luo2024experimental, he2023whistleblower, luo2025revisiting, tugrul2025understanding,Yuksel2025ColumnDisturb} extensively characterize the read disturbance in real DRAM chips (i.e., DDR3, DDR4, LPDDR4, and HBM2 chips). None of these works analyzes how the DRAM row initialization sequence and the \emph{previously} written data pattern affect the read disturbance vulnerability in real DRAM chips.

\noindent\textbf{PUD Operations in Real DRAM Chips.}
Several prior works demonstrate bulk bitwise~\cite{seshadri2017ambit,seshadri2015fast,hajinazar2021simdram,deoliveira2024mimdram,oliveira2025proteus} (e.g., AND, OR, and MAJority operations) and bulk data copy operations~\cite{seshadri2013rowclone} in real DRAM chips using multiple-row activation~\cite{olgun2021quactrng,gao2019computedram,gao2022frac,yaglikci2022hira,olgun2023dram,yuksel2024functionally,olgun2021pidram, mutlu2024memory}. Two of these prior works~\cite{gao2022frac,yuksel2024simultaneous} demonstrate that the reliability of PUD operation can be improved by replicating input operands~\cite{yuksel2024functionally} or storing fractional values in DRAM cells~\cite{gao2022frac}. No prior work analyzes or demonstrates how \emph{previously} written data to DRAM cells before the PUD operation can improve the reliability of such operations.

%% file: sections/09-conclusion.tex
\section{Conclusion}
\arxivaccept{This paper provides} the first experimental demonstration of DejaVu, a phenomenon where the data \emph{previously} written to DRAM cells affects DRAM's vulnerability to read disturbance\arxivaccept{, as well as data retention failures and the reliability of Processing-Using-DRAM (PUD) operations}. Our comprehensive experimental characterization shows that, compared to the baseline where we initialize the victim row by writing to it only once, 1) overwriting the victim row reduces \acmin (the minimum aggressor row activation count to induce at least one bitflip), and 2) writing the \emph{same} data to the victim row increases \acmin. We also find that overwriting the tested row with \arxivaccept{data \texttt{0xFF}} increases the number of DRAM retention failure bitflips compared to writing \arxivaccept{data \texttt{0xFF}} to the tested row twice. We provide two hypotheses to explain DejaVu and conduct controlled experimental characterization to provide more insights into the root cause(s) of DejaVu. Surprisingly, we also find that DejaVu improves the reliability of Processing-Using-DRAM (PUD) operations. Based on our observations, we discuss the implications of DejaVu on DRAM read disturbance testing and characterization methodology. We also evaluate the additional performance overhead of read disturbance mitigation techniques \arxivaccept{(e.g., PARA and PRAC)} when their read disturbance thresholds need to be lowered to be secure against DejaVu. We hope future works leverage our results, findings, and insights to build a more comprehensive and fundamental understanding of DRAM read disturbance\arxivaccept{, data retention, and Processing-Using-DRAM (PUD) operations} to build \arxivaccept{more robust and efficient DRAM-based computing systems.}

%% file: sections/10-artifact-appendix.tex
\appendix
\input{sections/extended_table}

\section{Artifact Appendix}
\subsection{Abstract}
The artifact contains the code and scripts to 1) run real DRAM chip characterization experiments on our specialized hardware infrastructure (Section IV, V, VII, and VIII), 2) evaluate the performance overhead of DRAM read disturbance mitigation techniques on a DRAM simulator (\secref{sec:mitigation}), and 3) parse, analyze, and plot the results.

The real DRAM chip characterization experiments require Xilinx Alveo U200 FPGA boards programmed with DRAM Bender~\cite{olgun2023dram, safari-drambender} and miscellaneous supporting hardware such as temperature controllers, heater pads, etc. \arxivaccept{The public release of our artifact does not contain the code to control the miscellaneous supporting hardware because they are proprietary to our internal infrastructure. We leave comments in the scripts to indicate where and how these hardware components should be controlled (according to the user's own hardware setup).} The simulator-based performance evaluation can run on any mainstream PC/server. We provide a Docker container image with all the software dependencies (e.g., python, c++-20 compiler toolchain, pip packages like pandas, matplotlib) already installed.

The artifact can reproduce all our key characterization results (Figure 1, 6-11, 13-22) and simulation results (Figure 23, 24). 

\subsection{Artifact check-list (meta-information)}

{\small
\begin{itemize}
  \item {\bf Hardware: }Xilinx Alveo U200 FPGA board, temperature controller and heater pads for DRAM modules under test, mainstream PC/server.
  \item {\bf Experiments: }Experimental real-DRAM chip characterization of DejaVu (read disturbance, retention failure, PUD operation reliability), and performance overhead evaluation for DRAM read disturbance mitigation techniques that need to take DejaVu into account. 
  \item {\bf Output: }Figure 1, 6-11, 13-24.
  \item {\bf How much disk space required (approximately)?:} 50GB
  \item {\bf How much time is needed to prepare workflow (approximately)?: } 1 day.
  \item {\bf How much time is needed to complete experiments (approximately)?: } ~20 days 
  \item {\bf Publicly available: }Yes.
  \item {\bf Code licenses: }MIT.
  \item {\bf Archived (DOI):} 10.5281/zenodo.19444878.
\end{itemize}
}

\subsection{Description}

\subsubsection{How to access}

The artifact can be accessed on Zenodo with DOI 10.5281/zenodo.19444878.

\subsubsection{Hardware dependencies}
Experimental characterization of real DRAM chips:
\begin{itemize}
\item Xilinx Alveo U200 FPGA board programmed with DRAM Bender
\item Temperature controller and heater pads for DRAM chips under test
\end{itemize}
Simulator-based performance overhead evaluation:
\begin{itemize}
\item Mainstream-spec x86 PC/server.
\end{itemize}
\subsubsection{Software dependencies}
Experimental characterization of real DRAM chips: Please check the README of DRAM Bender~\cite{safari-drambender} for software dependencies and installation instructions.

Simulator-based performance overhead evaluation: Docker
\subsection{Installation}

\noindent\textbf{Experimental characterization of real DRAM chips.} \arxivaccept{Follow the instructions in the README of DRAM Bender~\cite{safari-drambender}.}

\noindent\textbf{Simulator-based Performance Overhead Evaluation.} No installation is required as a Docker container image with all dependencies already installed will be provided.

\subsection{Evaluation and expected results}

\noindent\textbf{Experimental characterization of real DRAM chips.} 
\begin{enumerate}
    \item Extract the artifact zip file and \texttt{cd} into the extracted directory
    \item \texttt{cd} into \texttt{dejavu\_ae}
    \item Execute \texttt{run\_rd\_characterization.sh} to start the experimental characterization of DejaVu on DRAM read disturbance and retention failure on real DRAM chips
    \item Execute all cells in \texttt{dejavu\_rd\_plots.ipynb} to generate Figures 1, 6-11, and 13-21 in the notebook
    \item \texttt{cd} into \texttt{dejavu\_pud}
    \item Execute \texttt{run\_pud\_reliability.sh} to start the experimental characterization of DejaVu on improving PUD operations reliability on real DRAM chips
    \item Execute all cells in \texttt{analysis/plot\_pud.ipynb} to generate Figure 22 in the notebook
    
\end{enumerate}
\noindent\textbf{Simulator-based Performance Overhead Evaluation.}
\begin{enumerate}
    \item Start a docker container with the Dockerhub image \texttt{richardluo831/ramulator2}
    \item Copy the \texttt{perf\_eval} directory from the artifact into the container and \texttt{cd} into it
    \item Execute \texttt{run\_artifact.sh} to run all simulations
    \item Execute \texttt{parse\_results.sh} to collect and parse results
    \item Execute \texttt{plot\_all\_figures.sh} to generate Figure 23 and 24 at \texttt{ae\_results/dejavu/\_plots/}
\end{enumerate}

\subsection{Methodology}

Submission, reviewing and badging methodology:

\begin{itemize}
  \item \url{https://www.acm.org/publications/policies/artifact-review-and-badging-current}
  \item \url{https://cTuning.org/ae}
\end{itemize}

%% file: sections/extended_table.tex
\onecolumn
\newcommand{\AppendixLandscapePageDecor}{%
  \AddToHookNext{shipout/foreground}{%
    \begin{tikzpicture}[remember picture,overlay]
      \node[rotate=90] at ([xshift=-0.35in]current page.east) {\thepage};
    \end{tikzpicture}%
  }%
}
\pagestyle{empty}
\begin{landscape}
\AppendixLandscapePageDecor
\thispagestyle{empty}
\section{Tested DRAM Module Details}
\begin{table}[!ht]
\centering
\small
\setlength{\tabcolsep}{3pt}
\caption{Detailed information for the tested DDR4 DRAM modules; Average Double-Sided RowHammer \acmin at $50^\circ\mathrm{C}$ for 0$\rightarrow$1 (1$\rightarrow$0) bitflips for all tested modules, across 128 tested victim rows for each module, 50 repetitions, for the Baseline pattern where we write to the victim row only once, the SameWrite pattern where we initialize the victim row by writing the same data to it twice, and the OverWrite pattern where we initialize the victim row by first writing the opposite data, and then overwriting it with the actual victim data pattern. The percentage shows the change in \acmin of the SameWrite and OverWrite patterns relative to the Baseline pattern.}
\label{tab:dram-module-avg-acmin}
\begin{adjustbox}{max width=\linewidth}
\begin{tabular}{@{}c|c|cc|ccccc|ccc@{}}
\toprule
\multirow{2}{*}{\begin{tabular}[c]{@{}c@{}}\textbf{DRAM}\\\textbf{Manufacturer}\end{tabular}} &
\multirow{2}{*}{\textbf{ID}} &
\multirow{2}{*}{\textbf{DIMM Vendor}} &
\multirow{2}{*}{\textbf{DIMM Part Number}} &
\multirow{2}{*}{\textbf{DRAM Part Number}} &
\multirow{2}{*}{\begin{tabular}[c]{@{}c@{}}\textbf{Die}\\\textbf{Revision}\end{tabular}} &
\multirow{2}{*}{\begin{tabular}[c]{@{}c@{}}\textbf{Die}\\\textbf{Density}\end{tabular}} &
\multirow{2}{*}{\textbf{DQ}} &
\multirow{2}{*}{\textbf{Datecode}} &
\multicolumn{3}{c@{}}{\begin{tabular}[c]{@{}c@{}}\textbf{Average Double-Sided RowHammer \acmin;}\\\textbf{Bitflip direction 0$\rightarrow$1 (1$\rightarrow$0)}\end{tabular}} \\
& & & & & & & & &
\makebox[7em][c]{\textbf{Baseline}} &
\makebox[7em][c]{\textbf{SameWrite}} &
\makebox[7em][c]{\textbf{OverWrite}} \\
\midrule
\rowcolor{black!5}
\cellcolor{white} & S0 & Samsung & M378A1K43DB2-CTD & K4A8G085WD-BCTD & D & 8 Gb  & x8 & 2110 & 14165 (17469) & 14441; +1.9\% (17913; +2.5\%) & 13696; -3.3\% (16733; -4.2\%) \\
& S1 & Samsung & M378A4G43MB1-CTD & K4AAG085WW-BCTD & M & 16 Gb & x8 & N/A  & 16967 (19706) & 17428; +2.7\% (20327; +3.2\%) & 16443; -3.1\% (18855; -4.3\%) \\
\rowcolor{black!5}
\cellcolor{white} & S2 & Samsung & M378A2G43AB3-CWE & K4AAG085WA-BCWE & A & 16 Gb & x8 & 2302 & 16996 (20334) & 17431; +2.6\% (21036; +3.5\%) & 16411; -3.4\% (19342; -4.9\%) \\
& S3 & Samsung & M391A2G43BB2-CWE & K4AAG085WB-BCWE & B & 16 Gb & x8 & 2315 & 14077 (16747) & 14412; +2.4\% (17103; +2.1\%) & 13695; -2.7\% (16024; -4.3\%) \\
\rowcolor{black!5}
\cellcolor{white}\multirow{-5}{*}{Samsung} & S4 & Samsung & M471A4G43CB1-CWE & K4AAG085WC-BCWE & C & 16 Gb & x8 & 2408 & 10520 (12642) & 10718; +1.9\% (12903; +2.1\%) & 10260; -2.5\% (12154; -3.9\%) \\
\midrule
& H0 & SK Hynix & HMA81GU7AFR8N-UH & H5AN8G8NAFR-UHC & A & 8 Gb  & x8 & 1843 & 53980 (78572) & 55293; +2.4\% (80509; +2.5\%) & 52372; -3.0\% (75714; -3.6\%) \\
\rowcolor{black!5}
\cellcolor{white} & H1 & SK Hynix & HMA81GU6CJR8N-VK & H5AN8G8NCJR-VKC & C & 8 Gb  & x8 & 2120 & 25242 (37113) & 26057; +3.2\% (38461; +3.6\%) & 24183; -4.2\% (35888; -3.3\%) \\
& H2 & SK Hynix & HMA81GU7DJR8N-WM & H5AN8G8NDJR-WMC & D & 8 Gb  & x8 & 1938 & 20576 (31283) & 21421; +4.1\% (32081; +2.6\%) & 19902; -3.3\% (30102; -3.8\%) \\
\rowcolor{black!5}
\cellcolor{white}\multirow{-4}{*}{Hynix} & H3 & SK Hynix & HMAA4GU6AJR8N-VK & H5ANAG8NAJR-VKC & A & 16 Gb & x8 & 2003 & 28669 (42437) & 29448; +2.7\% (43701; +3.0\%) & 27734; -3.3\% (40975; -3.4\%) \\
\midrule
& M0 & Crucial  & CT16G4DFD824A.M16FE     & MT40A1G8SA-075:E   & E & 8 Gb  & x8 & 2402 & 51052 (52070) & 52287; +2.4\% (53511; +2.8\%) & 49236; -3.6\% (50276; -3.4\%) \\
\rowcolor{black!5}
\cellcolor{white} & M1 & Kingston & KSM32ES8/8MR            & MT40A1G8SA-062E:R  & R & 8 Gb  & x8 & 2412 & 22212 (26201) & 22701; +2.2\% (27235; +3.9\%) & 21476; -3.3\% (25118; -4.1\%) \\
& M2 & Kingston & KSM32ES8/16MF           & MT40A2G8SA-062E:F  & F & 16 Gb & x8 & 2412 & 15839 (17023) & 16232; +2.5\% (17615; +3.5\%) & 15440; -2.5\% (16513; -3.0\%) \\
\rowcolor{black!5}
\cellcolor{white} & M3 & Micron   & MTA18ASF4G72HZ-3G2F1Z1  & MT40A2G8SA-062E:F  & F & 16 Gb & x8 & 2237 & 16815 (18810) & 17147; +2.0\% (19377; +3.0\%) & 16355; -2.7\% (17988; -4.4\%) \\
\multirow{-5}{*}{Micron} & M4 & Micron   & MTA9ASF2G72AZ-3G2F1Z1   & MT40A2G8SA-062E:F  & F & 16 Gb & x8 & N/A  & 17601 (18248) & 18049; +2.5\% (18914; +3.6\%) & 17022; -3.3\% (17560; -3.8\%) \\
\bottomrule
\end{tabular}
\end{adjustbox}

\vspace{1.2em}
\setlength{\tabcolsep}{3pt}
\caption{Detailed information for the tested DDR4 DRAM modules; Minimum Double-Sided RowHammer \acmin at $50^\circ\mathrm{C}$ for 0$\rightarrow$1 (1$\rightarrow$0) bitflips for all tested modules, across 128 tested victim rows for each module, 50 repetitions, for the Baseline pattern where we write to the victim row only once, the SameWrite pattern where we initialize the victim row by writing the same data to it twice, and the OverWrite pattern where we initialize the victim row by first writing the opposite data, and then overwriting it with the actual victim data pattern. The percentage shows the change in \acmin of the SameWrite and OverWrite patterns relative to the Baseline pattern.}
\label{tab:dram-module-details}
\begin{adjustbox}{max width=\linewidth}
\begin{tabular}{@{}c|c|cc|ccccc|ccc@{}}
\toprule
\multirow{2}{*}{\begin{tabular}[c]{@{}c@{}}\textbf{DRAM}\\\textbf{Manufacturer}\end{tabular}} &
\multirow{2}{*}{\textbf{ID}} &
\multirow{2}{*}{\textbf{DIMM Vendor}} &
\multirow{2}{*}{\textbf{DIMM Part Number}} &
\multirow{2}{*}{\textbf{DRAM Part Number}} &
\multirow{2}{*}{\begin{tabular}[c]{@{}c@{}}\textbf{Die}\\\textbf{Revision}\end{tabular}} &
\multirow{2}{*}{\begin{tabular}[c]{@{}c@{}}\textbf{Die}\\\textbf{Density}\end{tabular}} &
\multirow{2}{*}{\textbf{DQ}} &
\multirow{2}{*}{\textbf{Datecode}} &
\multicolumn{3}{c@{}}{\begin{tabular}[c]{@{}c@{}}\textbf{Minimum Double-Sided RowHammer \acmin;}\\\textbf{Bitflip direction 0$\rightarrow$1 (1$\rightarrow$0)}\end{tabular}} \\
& & & & & & & & &
\makebox[7em][c]{\textbf{Baseline}} &
\makebox[7em][c]{\textbf{SameWrite}} &
\makebox[7em][c]{\textbf{OverWrite}} \\
\midrule
\rowcolor{black!5}
\cellcolor{white} & S0 & Samsung & M378A1K43DB2-CTD & K4A8G085WD-BCTD & D & 8 Gb  & x8 & 2110 & 8798 (10546) & 8898; +1.1\% (11205; +6.2\%) & 8496; -3.4\% (10216; -3.1\%) \\
& S1 & Samsung & M378A4G43MB1-CTD & K4AAG085WW-BCTD & M & 16 Gb & x8 & N/A  & 10536 (11434) & 10546; +0.1\% (11561; +1.1\%) & 10234; -2.9\% (10546; -7.8\%) \\
\rowcolor{black!5}
\cellcolor{white} & S2 & Samsung & M378A2G43AB3-CWE & K4AAG085WA-BCWE & A & 16 Gb & x8 & 2302 & 10546 (11718) & 10765; +2.1\% (12011; +2.5\%) & 10326; -2.1\% (11425; -2.5\%) \\
& S3 & Samsung & M391A2G43BB2-CWE & K4AAG085WB-BCWE & B & 16 Gb & x8 & 2315 & 9082 (11718) & 9082; 0.0\% (11718; 0.0\%) & 8789; -3.2\% (11369; -3.0\%) \\
\rowcolor{black!5}
\cellcolor{white}\multirow{-5}{*}{Samsung} & S4 & Samsung & M471A4G43CB1-CWE & K4AAG085WC-BCWE & C & 16 Gb & x8 & 2408 & 6518 (8203) & 6738; +3.4\% (8203; 0.0\%) & 6298; -3.4\% (7836; -4.5\%) \\
\midrule
& H0 & SK Hynix & HMA81GU7AFR8N-UH & H5AN8G8NAFR-UHC & A & 8 Gb  & x8 & 1843 & 25790 (44540) & 26604; +3.2\% (44531; 0.0\%) & 25195; -2.3\% (42773; -4.0\%) \\
\rowcolor{black!5}
\cellcolor{white} & H1 & SK Hynix & HMA81GU6CJR8N-VK & H5AN8G8NCJR-VKC & C & 8 Gb  & x8 & 2120 & 18750 (18750) & 19325; +3.1\% (22411; +19.5\%) & 17578; -6.3\% (18750; 0.0\%) \\
& H2 & SK Hynix & HMA81GU7DJR8N-WM & H5AN8G8NDJR-WMC & D & 8 Gb  & x8 & 1938 & 13924 (18750) & 14025; +0.7\% (19335; +3.1\%) & 13183; -5.3\% (18584; -0.9\%) \\
\rowcolor{black!5}
\cellcolor{white}\multirow{-4}{*}{Hynix} & H3 & SK Hynix & HMAA4GU6AJR8N-VK & H5ANAG8NAJR-VKC & A & 16 Gb & x8 & 2003 & 17578 (19335) & 18603; +5.8\% (20341; +5.2\%) & 17504; -0.4\% (19335; 0.0\%) \\
\midrule
& M0 & Crucial  & CT16G4DFD824A.M16FE     & MT40A1G8SA-075:E   & E & 8 Gb  & x8 & 2402 & 26962 (32235) & 27246; +1.1\% (32445; +0.7\%) & 25341; -6.0\% (28125; -12.8\%) \\
\rowcolor{black!5}
\cellcolor{white} & M1 & Kingston & KSM32ES8/8MR            & MT40A1G8SA-062E:R  & R & 8 Gb  & x8 & 2412 & 5273 (6774) & 5273; 0.0\% (7031; +3.8\%) & 4907; -6.9\% (6445; -4.9\%) \\
& M2 & Kingston & KSM32ES8/16MF           & MT40A2G8SA-062E:F  & F & 16 Gb & x8 & 2412 & 5859 (5859) & 6408; +9.4\% (6298; +7.5\%) & 5859; 0.0\% (5785; -1.3\%) \\
\rowcolor{black!5}
\cellcolor{white} & M3 & Micron   & MTA18ASF4G72HZ-3G2F1Z1  & MT40A2G8SA-062E:F  & F & 16 Gb & x8 & 2237 & 5566 (6481) & 5675; +2.0\% (6738; +4.0\%) & 5273; -5.3\% (6152; -5.1\%) \\
\multirow{-5}{*}{Micron} & M4 & Micron   & MTA9ASF2G72AZ-3G2F1Z1   & MT40A2G8SA-062E:F  & F & 16 Gb & x8 & N/A  & 7040 (7177) & 7644; +8.6\% (7617; +6.1\%) & 6884; -2.2\% (7031; -2.0\%) \\
\bottomrule
\end{tabular}
\end{adjustbox}
\end{table}

\clearpage
\AppendixLandscapePageDecor
\thispagestyle{empty}
\begin{table}[!ht]
\centering
\small
\setlength{\tabcolsep}{3pt}
\caption{Detailed information for the tested DDR4 DRAM modules; Average number of per-row retention failure bitflips for all tested modules, across 128 tested victim rows for each module, refresh disabled for 4096 ms at 95$^\circ$C, 50 repetitions, for the SameWrite pattern where we initialize the victim row by writing the same data to it twice, and the OverWrite pattern where we initialize the victim row by first writing the opposite data, and then overwriting it with the actual victim data pattern. The percentage shows the change in \acmin of the OverWrite patterns relative to the SameWrite pattern.}
\label{tab:retention-module-details}
\begin{adjustbox}{max width=\linewidth}
\begin{tabular}{@{}c|c|cc|ccccc|cc@{}}
\toprule
\multirow{2}{*}{\begin{tabular}[c]{@{}c@{}}\textbf{DRAM}\\\textbf{Manufacturer}\end{tabular}} &
\multirow{2}{*}{\textbf{ID}} &
\multirow{2}{*}{\textbf{DIMM Vendor}} &
\multirow{2}{*}{\textbf{DIMM Part Number}} &
\multirow{2}{*}{\textbf{DRAM Part Number}} &
\multirow{2}{*}{\begin{tabular}[c]{@{}c@{}}\textbf{Die}\\\textbf{Revision}\end{tabular}} &
\multirow{2}{*}{\begin{tabular}[c]{@{}c@{}}\textbf{Die}\\\textbf{Density}\end{tabular}} &
\multirow{2}{*}{\textbf{DQ}} &
\multirow{2}{*}{\textbf{Datecode}} &
\multicolumn{2}{c@{}}{\begin{tabular}[c]{@{}c@{}}\textbf{Average Number of}\\\textbf{Retention Failure Bitflips}\\\textbf{Per Row}\end{tabular}} \\
& & & & & & & & &
\makebox[6em][c]{\textbf{SameWrite}} &
\makebox[6em][c]{\textbf{OverWrite}} \\
\midrule
\rowcolor{black!5}
\cellcolor{white} & S0 & Samsung & M378A1K43DB2-CTD & K4A8G085WD-BCTD & D & 8 Gb  & x8 & 2110 & 256.0  & 280.8 (+9.7\%)  \\
& S1 & Samsung & M378A4G43MB1-CTD & K4AAG085WW-BCTD & M & 16 Gb & x8 & N/A  & 69.4   & 78.0 (+12.4\%)   \\
\rowcolor{black!5}
\cellcolor{white} & S2 & Samsung & M378A2G43AB3-CWE & K4AAG085WA-BCWE & A & 16 Gb & x8 & 2302 & 164.1  & 179.5 (+9.4\%)  \\
& S3 & Samsung & M391A2G43BB2-CWE & K4AAG085WB-BCWE & B & 16 Gb & x8 & 2315 & 318.5  & 344.3 (+8.1\%)  \\
\rowcolor{black!5}
\cellcolor{white}\multirow{-5}{*}{Samsung} & S4 & Samsung & M471A4G43CB1-CWE & K4AAG085WC-BCWE & C & 16 Gb & x8 & 2408 & 1138.5 & 1195.5 (+5.0\%) \\
\midrule
& H0 & SK Hynix & HMA81GU7AFR8N-UH & H5AN8G8NAFR-UHC & A & 8 Gb  & x8 & 1843 & 95.1  & 106.8 (+12.3\%) \\
\rowcolor{black!5}
\cellcolor{white} & H1 & SK Hynix & HMA81GU6CJR8N-VK & H5AN8G8NCJR-VKC & C & 8 Gb  & x8 & 2120 & 77.9  & 89.9 (+15.4\%)  \\
& H2 & SK Hynix & HMA81GU7DJR8N-WM & H5AN8G8NDJR-WMC & D & 8 Gb  & x8 & 1938 & 35.2  & 41.3 (+17.3\%)  \\
\rowcolor{black!5}
\cellcolor{white}\multirow{-4}{*}{Hynix} & H3 & SK Hynix & HMAA4GU6AJR8N-VK & H5ANAG8NAJR-VKC & A & 16 Gb & x8 & 2003 & 218.9 & 240.8 (+10.0\%) \\
\midrule
& M0 & Crucial  & CT16G4DFD824A.M16FE     & MT40A1G8SA-075:E   & E & 8 Gb  & x8 & 2402 & 557.9 & 610.0 (+9.3\%) \\
\rowcolor{black!5}
\cellcolor{white} & M1 & Kingston & KSM32ES8/8MR            & MT40A1G8SA-062E:R  & R & 8 Gb  & x8 & 2412 & 183.4 & 201.9 (+10.1\%) \\
& M2 & Kingston & KSM32ES8/16MF           & MT40A2G8SA-062E:F  & F & 16 Gb & x8 & 2412 & 445.1 & 471.7 (+6.0\%) \\
\rowcolor{black!5}
\cellcolor{white} & M3 & Micron   & MTA18ASF4G72HZ-3G2F1Z1  & MT40A2G8SA-062E:F  & F & 16 Gb & x8 & 2237 & 228.1 & 246.3 (+8.0\%) \\
\multirow{-5}{*}{Micron} & M4 & Micron   & MTA9ASF2G72AZ-3G2F1Z1   & MT40A2G8SA-062E:F  & F & 16 Gb & x8 & N/A  & 50.0  & 56.9 (+13.8\%)  \\
\bottomrule
\end{tabular}
\end{adjustbox}

\vspace{1.2em}
\setlength{\tabcolsep}{3pt}
\caption{Detailed information for the tested DDR4 DRAM modules; Maximum number of per-row retention failure bitflips for all tested modules, across 128 tested victim rows for each module, refresh disabled for 4096 ms at 95$^\circ$C, 50 repetitions, for the SameWrite pattern where we initialize the victim row by writing the same data to it twice, and the OverWrite pattern where we initialize the victim row by first writing the opposite data, and then overwriting it with the actual victim data pattern. The percentage shows the change in \acmin of the OverWrite patterns relative to the SameWrite pattern.}
\label{tab:retention-module-max-details}
\begin{adjustbox}{max width=\linewidth}
\begin{tabular}{@{}c|c|cc|ccccc|cc@{}}
\toprule
\multirow{2}{*}{\begin{tabular}[c]{@{}c@{}}\textbf{DRAM}\\\textbf{Manufacturer}\end{tabular}} &
\multirow{2}{*}{\textbf{ID}} &
\multirow{2}{*}{\textbf{DIMM Vendor}} &
\multirow{2}{*}{\textbf{DIMM Part Number}} &
\multirow{2}{*}{\textbf{DRAM Part Number}} &
\multirow{2}{*}{\begin{tabular}[c]{@{}c@{}}\textbf{Die}\\\textbf{Revision}\end{tabular}} &
\multirow{2}{*}{\begin{tabular}[c]{@{}c@{}}\textbf{Die}\\\textbf{Density}\end{tabular}} &
\multirow{2}{*}{\textbf{DQ}} &
\multirow{2}{*}{\textbf{Datecode}} &
\multicolumn{2}{c@{}}{\begin{tabular}[c]{@{}c@{}}\textbf{Maximum Number of}\\\textbf{Retention Failure Bitflips}\\\textbf{Per Row}\end{tabular}} \\
& & & & & & & & &
\makebox[6em][c]{\textbf{SameWrite}} &
\makebox[6em][c]{\textbf{OverWrite}} \\
\midrule
\rowcolor{black!5}
\cellcolor{white} & S0 & Samsung & M378A1K43DB2-CTD & K4A8G085WD-BCTD & D & 8 Gb  & x8 & 2110 & 322.0  & 352.4 (+9.4\%)  \\
& S1 & Samsung & M378A4G43MB1-CTD & K4AAG085WW-BCTD & M & 16 Gb & x8 & N/A  & 90.6   & 102.4 (+13.0\%) \\
\rowcolor{black!5}
\cellcolor{white} & S2 & Samsung & M378A2G43AB3-CWE & K4AAG085WA-BCWE & A & 16 Gb & x8 & 2302 & 193.7  & 214.4 (+10.7\%) \\
& S3 & Samsung & M391A2G43BB2-CWE & K4AAG085WB-BCWE & B & 16 Gb & x8 & 2315 & 387.0  & 414.5 (+7.1\%)  \\
\rowcolor{black!5}
\cellcolor{white}\multirow{-5}{*}{Samsung} & S4 & Samsung & M471A4G43CB1-CWE & K4AAG085WC-BCWE & C & 16 Gb & x8 & 2408 & 1255.6 & 1317.2 (+4.9\%) \\
\midrule
& H0 & SK Hynix & HMA81GU7AFR8N-UH & H5AN8G8NAFR-UHC & A & 8 Gb  & x8 & 1843 & 123.2 & 141.7 (+15.0\%) \\
\rowcolor{black!5}
\cellcolor{white} & H1 & SK Hynix & HMA81GU6CJR8N-VK & H5AN8G8NCJR-VKC & C & 8 Gb  & x8 & 2120 & 105.7 & 121.8 (+15.2\%) \\
& H2 & SK Hynix & HMA81GU7DJR8N-WM & H5AN8G8NDJR-WMC & D & 8 Gb  & x8 & 1938 & 56.7  & 63.2 (+11.5\%)  \\
\rowcolor{black!5}
\cellcolor{white}\multirow{-4}{*}{Hynix} & H3 & SK Hynix & HMAA4GU6AJR8N-VK & H5ANAG8NAJR-VKC & A & 16 Gb & x8 & 2003 & 259.4 & 286.7 (+10.5\%) \\
\midrule
& M0 & Crucial  & CT16G4DFD824A.M16FE     & MT40A1G8SA-075:E   & E & 8 Gb  & x8 & 2402 & 668.5 & 743.2 (+11.2\%) \\
\rowcolor{black!5}
\cellcolor{white} & M1 & Kingston & KSM32ES8/8MR            & MT40A1G8SA-062E:R  & R & 8 Gb  & x8 & 2412 & 221.7 & 244.1 (+10.1\%) \\
& M2 & Kingston & KSM32ES8/16MF           & MT40A2G8SA-062E:F  & F & 16 Gb & x8 & 2412 & 518.8 & 551.3 (+6.3\%)  \\
\rowcolor{black!5}
\cellcolor{white} & M3 & Micron   & MTA18ASF4G72HZ-3G2F1Z1  & MT40A2G8SA-062E:F  & F & 16 Gb & x8 & 2237 & 275.5 & 293.8 (+6.6\%)  \\
\multirow{-5}{*}{Micron} & M4 & Micron   & MTA9ASF2G72AZ-3G2F1Z1   & MT40A2G8SA-062E:F  & F & 16 Gb & x8 & N/A  & 72.1  & 81.4 (+12.9\%)  \\
\bottomrule
\end{tabular}
\end{adjustbox}
\end{table}
\end{landscape}

\clearpage
\pagestyle{fancy}
\twocolumn